\documentclass[11pt,a4paper]{article}
\usepackage{amsmath}
\usepackage{enumerate}
\usepackage{amssymb}
\usepackage{graphicx}
\usepackage{multirow}
\usepackage{xcolor}
\usepackage{tablefootnote}
\usepackage{threeparttable}
\usepackage{bm}
\usepackage[normalem]{ulem}
\usepackage{jcappub}

\newcommand{\Mag}{\mathcal{M}}
\begin{document}

\title{The Age of the Universe with Globular Clusters: reducing systematic uncertainties}
\author[a,b]{David Valcin,}
\author[a,c]{Raul Jimenez,}
\author[a,c]{Licia Verde,}
\author[d]{Jos\'e Luis Bernal,}
\author[e,f,g]{Benjamin D.~Wandelt}

\affiliation[a]{ICC, University of Barcelona, Mart\' i i Franqu\` es, 1, E08028
Barcelona, Spain}
\affiliation[b]{Dept. de F\' isica Qu\` antica i Astrof\' isica, University of Barcelona, Mart\' i i Franqu\` es 1, E08028 Barcelona,
Spain}
\affiliation[c]{ICREA, Pg. Lluis Companys 23, Barcelona, 08010, Spain.} 
\affiliation[d]{Department of Physics and Astronomy, Johns Hopkins University, 3400 North Charles Street, Baltimore, Maryland 21218, USA}
\affiliation[e]{Sorbonne Universit\'e, CNRS, UMR 7095, Institut d'Astrophysique de Paris, 98 bis bd Arago, 75014 Paris, France.}
\affiliation[f]{Sorbonne Universit\'e, Institut Lagrange de Paris (ILP), 98 bis bd Arago, 75014 Paris, France.}
\affiliation[g]{Center for Computational Astrophysics, Flatiron Institute, 162 5th Avenue, 10010, New York, NY, USA.}

\emailAdd{d.valcin@icc.ub.edu}
\emailAdd{jbernal2@jhu.edu} 
\emailAdd{raul.jimenez@icc.ub.edu}
\emailAdd{liciaverde@icc.ub.edu}
\emailAdd{bwandelt@iap.fr}

\abstract{The dominant systematic uncertainty in the age determination of galactic globular clusters is the depth of the convection envelope of the stars. This parameter is  partially degenerate with metallicity which is in turn degenerate with age. 
However, if  the metal content, distance and extinction are  known, the position and morphology  of the red giant branch in a color-magnitude diagram  are mostly sensitive to the value of the depth of the convective envelope.
Therefore, using external, precise metallicity determinations this degeneracy  and thus the systematic error in  age, can be reduced. 
Alternatively, the morphology of the  red giant branch of globular clusters color magnitude diagram can  also be used to achieve the same.   
We demonstrate that globular cluster red giant branches are well fitted by values of the  depth of the  convection envelope  consistent with those obtained for the Sun and this finding is robust to the adopted treatment of the stellar physics. With these findings, the uncertainty in the depth of the convection envelope  is no longer the dominant contribution to the systematic error in the age determination of the oldest globular clusters, reducing it  from $0.5$ to $0.23$ or $0.33$ Gyr, depending on the methodology adopted: i.e., whether resorting to external data (spectroscopic metallicity determinations) or relying solely on the morphology of the clusters's color-magnitude diagrams. This  results in an age of the Universe  $t_{\rm U}=13.5^{+0.16}_{-0.14} {\rm (stat.)} \pm 0.23(0.33) ({\rm sys.})$ at 68\% confidence level, accounting for the formation time of globular clusters and its uncertainty. An uncertainty of  0.27(0.36)  Gyr if added in quadrature. This agrees well with $13.8 \pm 0.02$ Gyr, the cosmological model-dependent value inferred by the Planck mission assuming the $\Lambda$CDM model.}

\maketitle

\section{Introduction}
\label{sec:intro}

A Bayesian analysis to estimate,  as precisely and accurately as possible, the absolute ages of galactic globular clusters (GCs) with resolved stellar populations was presented in a recent paper~\cite{ValcinGC}. The objective of the work in Ref.~\cite{ValcinGC} was to use the age of the oldest GCs to obtain an estimate of  the age of the Universe insensitive to cosmology
and, in turn, constrain cosmological models. By using the morphology of the color-magnitude diagram (CMD) and not just the luminosity of the main sequence turn off, we showed that the age, distance and metal content could be determined without relying on external data sets. By using the extensive set of GC CMDs from the ACS-HST survey, an age for the oldest GCs of $t_{\rm GC}=13.32 \pm 0.1 {\rm (stat.)} \pm 0.5 {\rm (sys.)}$, at 68\% confidence level, was obtained. As it is apparent, the uncertainty in the age is dominated by the systematic uncertainty, which in turn dominates the estimate of  the age of the Universe (see also~\cite{OMalley,BayesianGC,JimGC}).

The most important ``known unknown" contributing to  the systematic uncertainty  budget is the value of the depth of the convection envelope in low mass stars (those around solar mass). This by itself contributes to 60\% of the systematic uncertainty (see Table 2 in Ref.~\cite{OMalley}, the rest of the systematic error budget being due to reaction rates and opacities). The problem is at follows:  low mass stars have fully convective and turbulent envelopes (Reynolds number $\simeq 10^{10}$) and because of this, a full hydro-dynamical solution is prohibitive for a large grid of stellar models varying parameters like mass, metallicity and age (this can be done for a single star, and it is done when modelling  the Sun, but cannot -yet- be extended to a full library of stellar models). Instead, one models the  gradient of the convective transport by assuming 1D geometry and following a convective cell as it dissolves into the envelope. With this approach, the equations of stellar structure contain five independent differential equations for five variables and an extra parameter: the mixing length ($\alpha_{\rm MLT}$)\footnote{To be clear on nomencalture, hereafter   when we refer to mixing length or $\alpha_{\rm MLT}$, we mean the value adopted for the stellar interior in the convection theory,  and not the fixed value used for model atmosphere}. The value of the mixing length  parameter has to be obtained from fits to observations.  While there is some recent theoretical progress on matching 3D to 1D models for low mass stars (see e.g., Ref.~\cite{Weiss}) which could open the possibility to eliminate the need to empirically calibrate $\alpha_{\rm MLT}$,  this step cannot be avoided at the moment.

The standard way to determine the free mixing length parameter is to fit it to the Sun and assume this value applies to all stars. This, of course, is an assumption that is not guaranteed to hold for stars in GCs which have very different metallicity than the Sun.\footnote{See the discussion in section 2.2 in Ref.~\cite{OMalley} and references therein.}  While the adopted value for the mixing length  parameter does not affect the age determination directly, it indirectly does so  via  degeneracies with other parameters, most notably metallicity. The approach of Ref.~\cite{ValcinGC}  is to propagate a variation of the mixing length parameter over a wide range into the systematic error budget for the age, as adopted  and motivated by e.g., Ref.~\cite{OMalley}. 
 However,  as anticipated in Refs.~\cite{ValcinGC,JimenezGC96} this does not need to be the case as  the mixing length can, at least in principle, be constrained from the morphology of the CMD of GCs. We address this methodology in this article. 

\begin{figure}
\centering
\includegraphics[width=\textwidth]{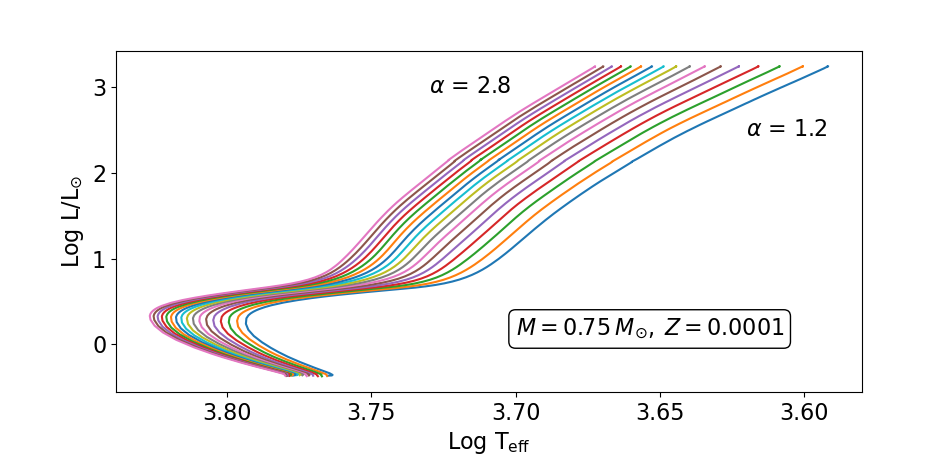} 
\caption{Variation of the HR diagram due to changes in the mixing length parameter $\alpha_{\rm MLT}$ ($\Delta_{\alpha}$ = 0.1) for a star with a fixed initial mass and metallicity. As we can see, the color (effective temperature) of the RGB is the most sensitive region to the mixing length value, while the sub-giant branch is the least sensitive part of the HR diagram.}
\label{fig:alpha}
\end{figure}

The rationale behind this approach is simple. As a first step, let us assume that the metallicity of the GC has been determined (for example, via spectroscopic observations). For a fixed metallicity, the color of the red giant branch (RGB) in a theoretical CMD depends mostly on the value of $\alpha_{\rm MLT}$ (see Fig.~\ref{fig:alpha}). As we will show below, other parameters affecting stellar structure do not modify the color of the RGB as much as $\alpha_{\rm MLT}$. Hence the  spread in color  of the RGB in the CMD of a single GC 
 yields an upper limit to the star-to-star variations in $\alpha_{\rm MLT}$ (assuming the scatter is solely due to  spread in $\alpha_{\rm MLT}$ values). 
Without resorting to external constraints on the GC metallicity, the metallicity determination for each GC of  Ref.~\cite{ValcinGC}, obtained assuming a fixed fiducial mixing length parameter value, should be affected by an (unknown) shift induced by the $\alpha_{\rm MLT}$ choice. Now, if the distance is known,  GCs of similar estimated metallicity can be suitably aligned on the theoretical CMD (or an HR diagram).  In this case, the dispersion in color of the RGB can be used to constrain the $\alpha_{\rm MLT}$ range, in particular if one assumes that the full scatter is solely induced by $\alpha_{\rm MLT}$.

In this paper we quantify this dispersion and constrain the range of  $\alpha_{\rm MLT}$  values. This significantly reduces the systematic uncertainty in the age estimation of GCs, making the mixing length contribution to the statistical error budget now subdominant to other systematics, and propagates into a determination of the age of the Universe with systematic errors reduced by $\sim$ 50\%.

\section{Data} 
\label{sec:data}
Following Ref.~\cite{ValcinGC}, we consider the globular clusters from the HST-ACS catalog which we group  into three metallicity samples (according to the best fit metallicity value): 12 clusters with [Fe/H] $<2.0$, 11 clusters with $-2<$ [Fe/H] $<$ -1.75 and 15 with  $-1.75<$ [Fe/H] $<$ -1.5.
One of the clusters in  Ref.~\cite{ValcinGC}, (NGC6715) shows clear signs of multiple populations in the RGB, its metallicity is just at the high edge of the range considered in this work and  its age determination has a very large error-bar. We exclude this cluster from the present analysis, leaving us with a sample of 38 clusters (including it does not change the results in any significant way, due to the large uncertainties on its age). 

\begin{figure}
\centering
\includegraphics[width=0.8 \textwidth]{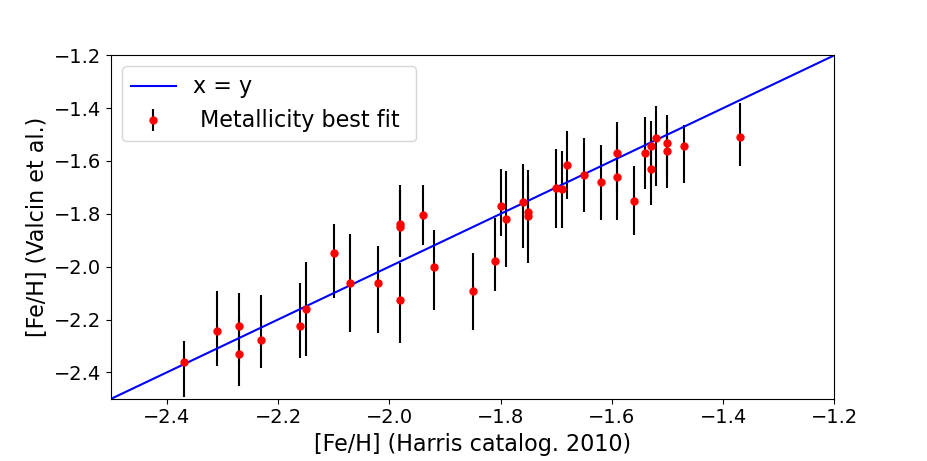} 
\caption{Metallicity determination of Ref.~\cite{ValcinGC} vs spectroscopic metallicity determination of Ref.~\cite{Harris2} of the 38 clusters in our sample. The 1:1 line guides the eye. The scatter  around this relation is $\sigma_{\rm [Fe/H]}=0.09$, with no indication of a dependence on metallicity or  systematic bias (i.e., a systematic deviation from the 1:1 line). }
\label{fig:specmetvsmet}
\end{figure}

These clusters also have spectroscopically-determined metallicities from Ref.~\cite{OMalley} (for only 16 of the 38 clusters) and \cite{Harris, Harris2} (for all 38);  the latter determination is complete for our purposes, then it is the main one we use here.  The comparison between spectroscopic metallicity and the metallicities estimated by  Ref.~\cite{ValcinGC} is shown in Fig.~\ref{fig:specmetvsmet}.  The comparison with Ref.~\cite{OMalley}  metallicities can be found in Fig. 5, of  Ref.~\cite{ValcinGC}.  

\begin{figure}
\centering
\includegraphics[width=0.8\textwidth]{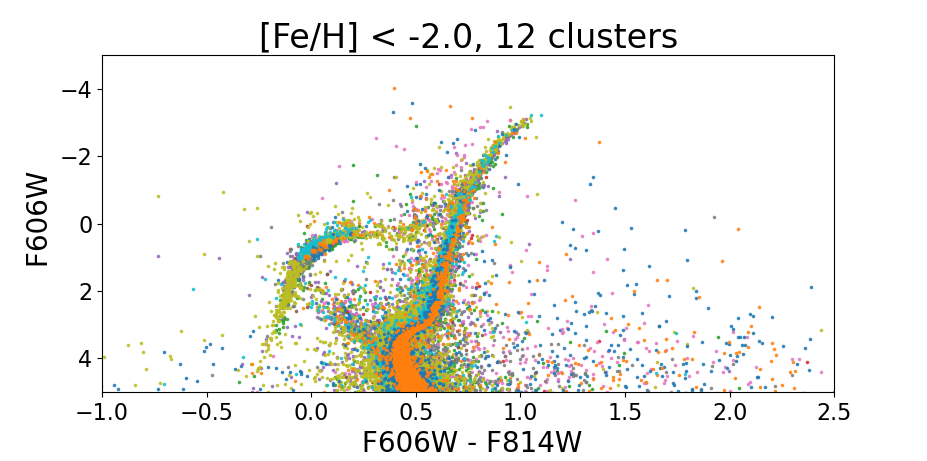} 
\includegraphics[width=0.8\textwidth]{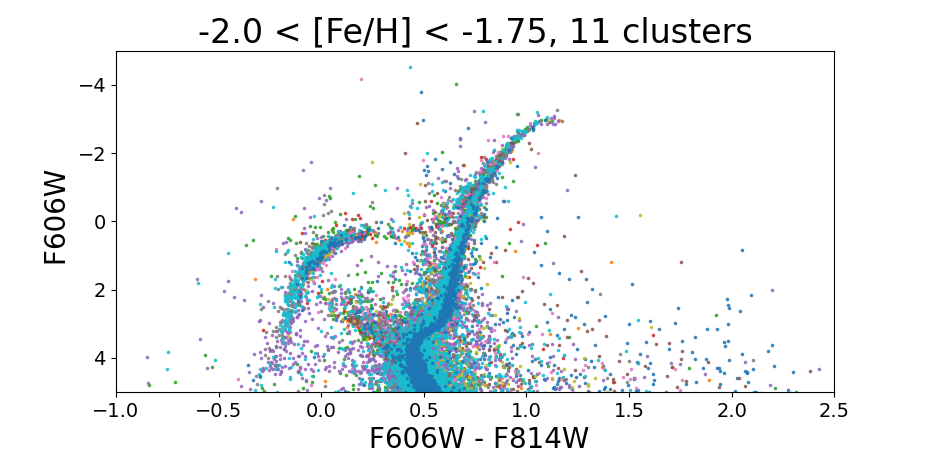} 
\includegraphics[width=0.8\textwidth]{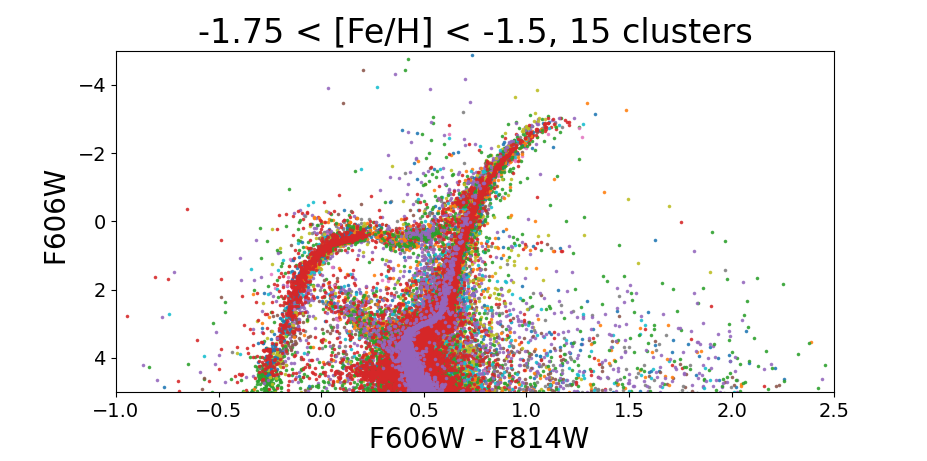} 
\caption{Top panel: Combined CMD of the  GCs with metallicities below [Fe/H] $< -2$, shifted to be at the same distance using the best fit distances and absorption from Ref.~\cite{ValcinGC}. Middle and bottom panels: same as top panel but for metallicity ranges as indicated. }
\label{fig:allGC}
\end{figure}

 In order to compare the clusters with each other and with the stellar tracks, it is necessary to convert the apparent magnitudes of each star into absolute magnitudes. To do this we use  the best fits obtained in Ref.~\cite{ValcinGC} (see their Table 3, Appendix E) for the absorption and the distance modulus. In the three panels of Fig.~\ref{fig:allGC},  we  show  the (absolute) CMD of all the clusters, as if they were all at the same distance (i.e., 10 pc), subdividing the sample in the three metallicity ranges listed above.  Clearly,  the RGBs of the different clusters appear nicely aligned in each metallicity interval. When considering the combined distribution of all the stars from all the GC in each sample,  there are many possible contributions to the resulting width of the RGB:  photometric errors,  errors in the best-fit parameter values (e.g., distance, absorption) used to  generate the plots,  errors in metallicity determinations and  the spread in metallicity within the selected sample, and the effects of variations in $\alpha_{\rm MLT}$ (which is the quantity we are interested in). 

\section{Method}
\label{sec:method}

Several parameters affect the color of the RGB in low-mass stars ($< 2 M_{\odot}$). The most important one is the metallicity content, as can be seen from  Fig. 8 in Ref.~\cite{ValcinGC}, where one can appreciate that the color of the RGB is quite insensitive to age, but  sensitive to metallicity. The  next  leading parameter determining the color of the RGB is the mixing length ($\alpha_{\rm MLT}$),  as we will show below by varying the parameters of the microphysics in the star and comparing the resulting stellar tracks (see Fig.~\ref{fig:alpha}). 
As recognized by Ref.~\cite{OMalley},  this parameter dominates the systematic uncertainty when obtaining ages of GCs using the luminosity of the main sequence turn off.

Here we explore two approaches to reduce this uncertainty. The first one is based on external metallicity determinations, the second one uses only internal information from the  morphology of the CMDs of each GC.

\subsection{Resorting to Spectroscopic metallicity determination}
\label{sec:methods0}
 We can appreciate in   Fig.~\ref{fig:specmetvsmet} that   there is a good agreement between spectroscopic and CMD-estimated metallicities, with no indication of a dependence on metallicity and  no indication of a  systematic bias (i.e., a systematic deviation from the 1:1 line). 
 When a linear fit is performed, the best-fit line  has a slope of $0.89\pm 0.15$ and an intercept of $-0.21\pm 0.28$.  When forcing the line to have a slope of $1$  a possible systematic normalization shift in the metallicity determination is $-0.023\pm 0.093$.  We find similar results when limiting this comparison to each of the subsamples in metallicity. Hence, we quantify the scatter in the relation i.e., a possible  difference between  spectroscopic and CMD-estimated metallicity to be  $\sigma_{\rm [Fe/H]}=0.093$.

  These results are obtained using the metallicities from Ref.~ \cite{Harris2};  results obtained using  Ref.~\cite{OMalley} instead are consistent,  but more uncertain  because of  the smaller number of objects included.   Since the metallicity determination of Ref.~\cite{ValcinGC} is obtained for a  value of $\alpha_{\rm MLT}=1.938$ fixed a priori,\footnote{The value of  the $\alpha_{\rm MLT}$ parameter assumed depends on the stellar code used. In Ref.~\cite{ValcinGC},  the values  reported were according to the convention of  the codes used there  i.e., \texttt{DSED}. Here we convert to  the convention of the  \texttt{MESA} (and  \texttt{JimMacD})  codes, see below. For reference the solar value for $\alpha_{\rm MLT}$ are  1.938, 2, 1.4, respectively. This change, however, only amounts to a shift; the relevant quantity  for our argument is the interval or range adopted, which does not depend on the convention.} and there is an expected degeneracy between  $\alpha_{\rm MLT}$ and metallicity,  a  systematically incorrect choice of  $\alpha_{\rm MLT}$ would have biased the metallicity determination, and hence the age. On the other hand, an  incorrect value of  $\alpha_{\rm MLT}$ with a  cluster-to-cluster variation would  induce a scatter in the comparison of Fig.~\ref{fig:specmetvsmet}, which we estimate to be $\sigma_{\rm [Fe/H]}=0.093$.  Below, we will estimate the allowed range of $\alpha_{\rm MLT}$  by attributing the full scatter of this relation to variations in the  mixing length parameter.  

\subsection{Using only internal information}
When using the full morphology of the CMD, it should be possible to  treat $\alpha_{\rm MLT}$ as an additional model parameter  to be constrained by the data, as mentioned in Ref.~\cite{ValcinGC}. 
 Here we develop this idea. We start by illustrating  the sensitivity of the RGB to stellar parameters and in particular to $\alpha_{\rm MLT}$ using   the publicly available  1D stellar structure and evolution  codes \texttt{MESA}~\cite{MESA} and the \texttt{JimMacD} code~\cite{Jimenez95}. These codes compute the 1D equations of stellar structures and evolve them in time, thus providing the structure of a star and its position and evolution in time  in the theoretical CMD  for  given initial  mass and chemical composition.  The numerical solution  of the stellar structure equations  of both codes are the same. The main  difference between the two codes is that 
\texttt{MESA} is a modern 1D stellar code that employs new updates in opacities and nuclear reaction rates. On the other hand, the older version of \texttt{JimMacD} that we use adopts different  values for opacities and nuclear reaction rates and a different formulation of the mixing length formulation.  We use these two different codes to illustrate that recent updates in nuclear reaction rates and opacities do not affect our results.
We then proceed to  constrain the mixing length parameter from the color of the RGB and quantify its spread for the oldest GCs.

\subsubsection{Color transformation}
\label{sec:colortransf}
The output of \texttt{MESA} allows us to plot directly the evolution of a star in the theoretical HR diagram (effective temperature vs log luminosity), but if we want to compare our tracks with the GCs observations we must transform the luminosity into the magnitudes corresponding to the filters of the HST-ACS catalog (F606W and F814W). 
The transformation is carried out in 4 stages:
\begin{enumerate}
\item convert luminosity to bolometric magnitude using the formula 
\[\rm \Mag_{bol} = -2.5 \:\rm log_{10} \:\frac{L_{\star}}{L_{0}}\] where $L_{\star}$  is the star's bolometric luminosity  in watts and $L_{0}$ is the zero point luminosity $= 3.0128 \times 10^{28} $ W\,,
\item produce bolometric correction tables using the Vega calibration (Calspec Alpha Lyrae)\footnote{https://ssb.stsci.edu/cdbs/current\_calspec/} and the atmospheric models of Castelli \& Kurucz \citep{Castelli} for various  metallicity values, 
 or the \texttt{Bolometric Corrections} code from Casagrande \& VandenBerg \cite{Casagrande}\footnote{For the rest of the paper we choose to work with bolometric corrections from Ref.~\cite{Casagrande}.}\,. This is further discussed in Appendix~\ref{sec:metmass}, where we show that the choice of  bolometric correction is unimportant for our purpose,
\item interpolate  bolometric corrections (BC) using effective temperature, surface gravity, and extinction where we assume  to have corrected the data for extinction and therefore take    E(B - V) = 0\,,
\item transform bolometric magnitude into absolute magnitude \[\rm \Mag_{filter} = \Mag_{bol} - BC_{filter}\] where $\rm \Mag_{filter}$ and $\rm BC_{filter}$ are respectively the absolute magnitude and the bolometric correction in the desired filter.
\end{enumerate}

\subsubsection{Stellar tracks}
\label{tracks}
The CMD of a GC is an isochrone which covers a range of initial masses for the stars. In the literature isochrones packages are usually computed from evolutionary tracks with different sets of parameters (See Table \ref{tab:param_iso} where we listed some of available parameters). One thing that all isochrones packages have in common is that they are all calibrated with respect to a fixed $\alpha_{\rm MLT}$ value. As we want to study the impact of this parameter on the position of the RGB in the CMD, either we must recompute a new grid of tracks with $\alpha_{\rm MLT}$ as an extra parameter to work with isochrones or we must work directly with tracks. We chose the latter since it is less expensive computationally. It is not an issue since it is well known that a stellar track for  a fixed mass, corresponding to the mass of the main sequence turn off, will approximate very well the parts of  CMD of the GC  which we consider here, in particular the upper part of the  RGB (where the dependence on the mass is very small). 

\begin{table}[h!]
\noindent\makebox[\textwidth][c]{%
\begin{threeparttable}{}
\begin{tabular}{|c|c|c|c|c|}
\hline
Stellar model & DSED & MIST & BASTI & PARSEC\\ \hline\hline
Mixing length  $\alpha_{MLT}$ & 1.93 & 1.82 & 1.913 & 1.74\\ \hline
Age range (Gyr) & 0.250-15 & 0.0001-20 & 0.0226-19.1972\tnote{\ddag} & 0.001-20 \\ \hline
Metallicity range {[}Fe/H{]} & -2.5 to 0.5 dex & -4.0 to 0.5 dex & -3.27 to 0.51 dex\tnote{\ddag} & -2.2 to 0.5 dex \\ \hline
Initial Helium fraction $Y_{\rm init}$ & 0.245, 0.33\tnote{\dag}, 0.40\tnote{\dag} & 0.249 & 0.245 & 0.2485\\ \hline
Helium fraction configuration &  $Y = Y_{\rm init} + 1.5\times Z$ &  $Y = Y_{\rm init} + 1.5\times Z$ &  $Y = Y_{\rm init} + 1.4\times Z$ & $Y = Y_{\rm init} + 1.78\times Z$ \\ \hline
 {[}$\alpha$/Fe{]} & -0.2 to 0.8 & N/A & 0.4 & N/A\\ \hline
 \end{tabular}%
\footnotesize
\begin{tablenotes}
\item[\dag] Fixed Helium fraction configurations $Y=0.33 $ and 0.40 are only available for [Fe/H] $\leq$ 0.
\item[\ddag] Values for the solar-scaled model.
\end{tablenotes}
\end{threeparttable}}
\caption{Properties of stellar models available in the literature.}
\label{tab:param_iso}
\end{table}

 The RGB for an isochrone  of a fixed age can always be approximated by a stellar track for a suitable choice of the initial mass. If we compare an isochrone of fixed age to a track of similar mass, we see that the RGB phase lasts much less time (between a factor 20 and 100) than the main sequence and therefore the variation in mass on the RGB isochrone is small which justifies our approximation (see table  \ref{tab:lifetime} for a summary). To define the different phases, we used \texttt{MIST} evolutionary points EEPs 202, 454 and 605 corresponding respectively to the zero age main sequence, terminal age main sequence and the tip of the red giant branch. We fix the age for the isochrone to be 13.32 Gyr and find that $ M = 0.75\:M_{\odot}$ yields the best fit for our purposes.  A variation of the value of  the initial mass  has an effect on an isochrone  very similar to changing the age,  and  mainly affects the main sequence and its turn off.. This is further discussed in  Appendix \ref{sec:metmass}, especially  Fig. \ref{fig:mass}, but   see also  Figure 8 in Ref.~\cite{ValcinGC}.
 
\begin{table}[h!]
    \centering
    \begin{tabular}{|c|c|c|c|}
    \hline
    Type & Fixed quantity & $\Delta_{MS}$ & $\Delta_{RGB}$ \\ \hline \hline
       Track  & Mass & $14.98$ Gyr & $0.86$ Gyr\\ \hline
        Isochrone  &  Age & $0.66 \: M_{\odot}$ & $0.01 \: M_{\odot}$ \\ \hline
    \end{tabular}
    \caption{Difference of lifetime and mass between the main sequence and the RGB. The value are given for an isochrone of Age = 13.32 Gyr and a track of mass $M = 0.75 M_{\odot}$ for [Fe/H] $= -2.0$.}
    \label{tab:lifetime}
\end{table}
 
\begin{figure}[h!]
    \centering
    \includegraphics[width=0.495\textwidth]{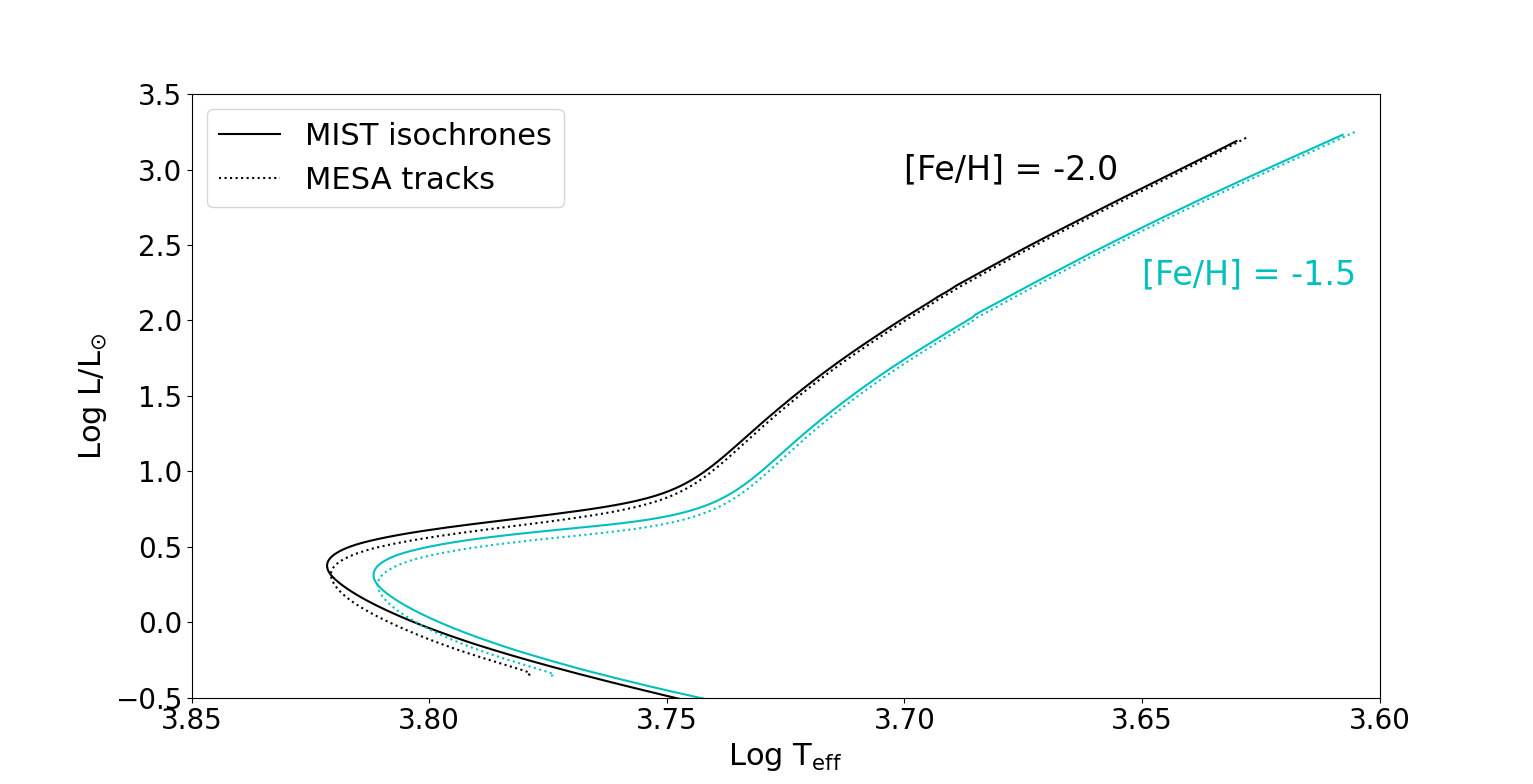}
    \includegraphics[width=0.495\textwidth]{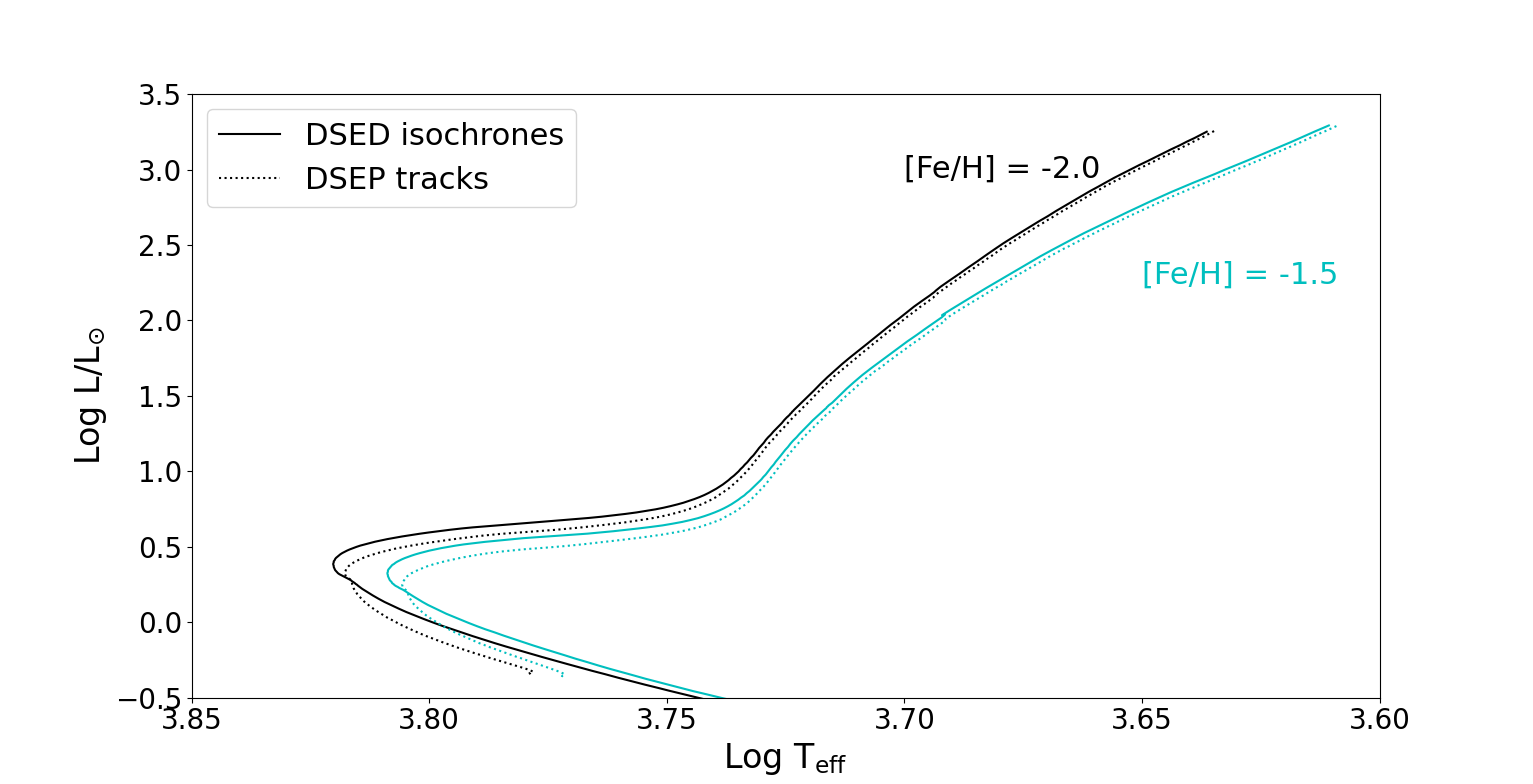}
    \includegraphics[width=0.495\textwidth]{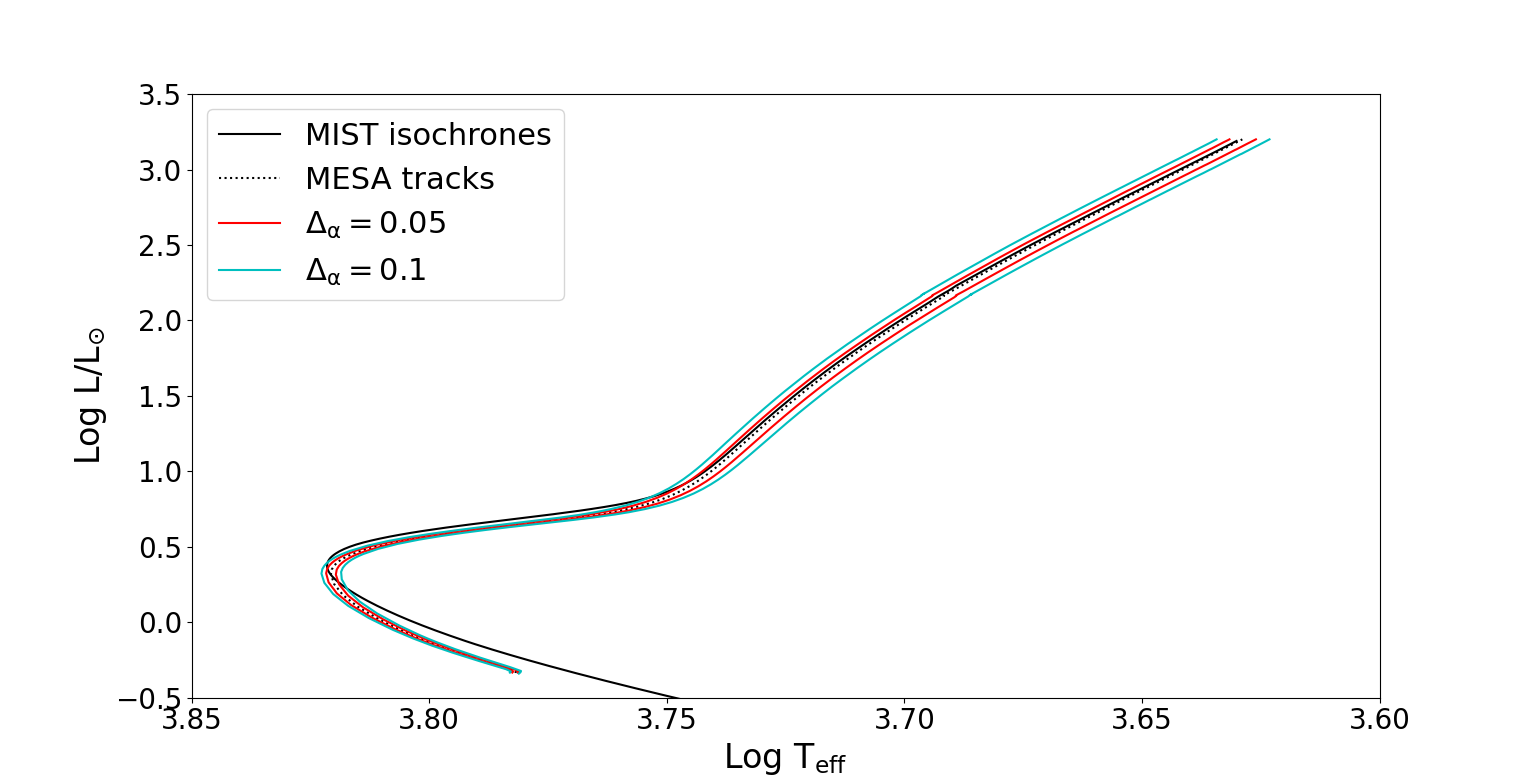}
    \includegraphics[width=0.495\textwidth]{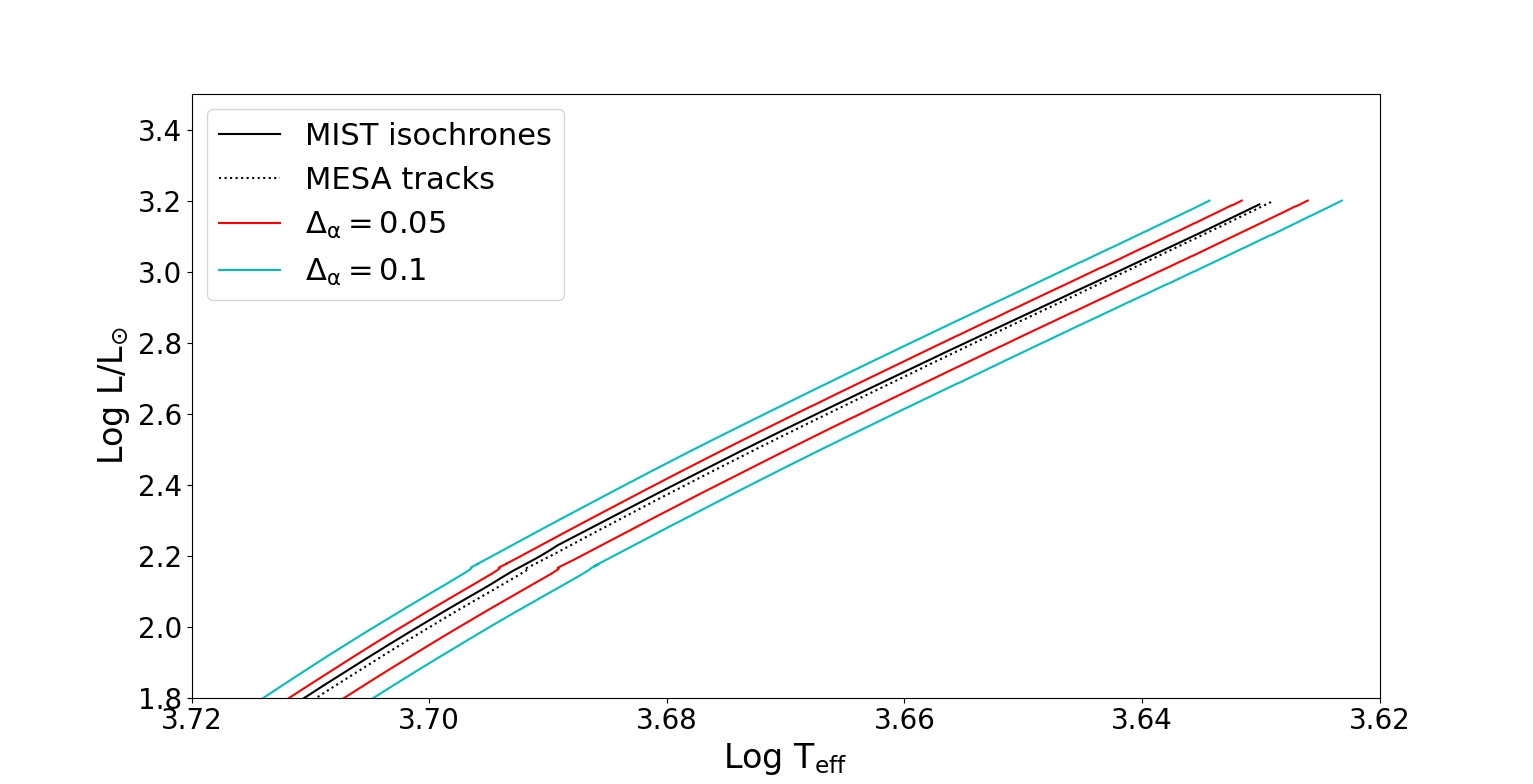}
    \caption{The upper part of the RGB of an isochrone is very well approximated by the stellar track for a suitable choice of mass (and age, but the sensitivity to age is small). For each isochrone the age is equal to 13.32 Gyr and for each track the mass is equal to $0.75M_{\odot}$. In this figure the dotted lines correspond to isochrones, the solid lines to stellar tracks. The top panel shows the agreement for two representative values of metallicity of stellar models couples: (left) \texttt{MIST} vs \texttt{MESA} and (right) \texttt{DSEP} vs \texttt{DSEP}; the bottom panel shows that the differences between isochrones and tracks are much smaller than those induced by a change in $\alpha_{\rm MLT}$ of the magnitude of interest.
}
    \label{fig:isotrack}
\end{figure}

 The isochrone-track agreement  is illustrated in Fig.~\ref{fig:isotrack}. In both panels, the dotted lines correspond to isochrones, and the solid lines, to stellar tracks  for representative values of metallicity and mixing length parameter. For the magnitude range ($\Mag \gtrsim 0$), and the  combination of parameters  we are interested in,  the differences between  isochrones and stellar tracks are very small and completely negligible compared to the differences induced by  changes in metallicity and $\alpha_{\rm MLT}$ considered here. In our analysis, we also include a cut for magnitudes $\Mag \gtrsim -2$; for $\Mag<0$ the differences between isochrones and tracks as obtained according to our procedure of sec.~\ref{sec:colortransf}, are slightly more pronounced. For this reason, the $\Mag>-2$ cut  is only reported in the appendices and serve to check for possible effects of outliers and to cap the effect of such mismatch.

%
%

We begin by  computing stellar tracks using the \texttt{MESA} software package~\cite{MESA}, evolving a star from pre main-sequence to a luminosity limit of $\rm log_{10} (L_{max}$) = 3.25, sufficient to compare the tip of the RGB for different values of the mixing length parameter. Among the various parameters needed to configure the tracks, the initial mass and metallicity are the two most important. 

We calculate the stellar tracks for  8  initial values of metallicity spanning a range  from $Z = 0.00005$ to $Z = 0.0004$ (equivalent to $\rm -2.45 < [Fe/H] < -1.55$)\footnote{Recall that [Fe/H] $=  \log_{10}(Z/Z_{\odot})$ where $Z_{\odot}=0.02$ for \texttt{JimMacD} and  $Z_{\odot}=0.0142$ for  \texttt{MESA} .} to sample  the $\rm [Fe/H]$ range of  the 38 GCs in our sample  (see Table 3 in Ref.~\cite{ValcinGC}). Except for its tip, the RGB varies only slightly with changes in the metallicity. 
 This is why we argue that, for a metallicity range  comparable to current uncertainties from CMD studies, the scatter around the the RGB may be used to constrain the maximum  range of $\alpha_{\rm MLT}$. The  impact of the choice of mass, metallicity and the stellar model on the stellar tracks and the CMD is explored in more detail in Appendix~\ref{sec:metmass} (see Figure \ref{fig:metal}).

Besides adopting the  solar abundance scale, we use the same configuration parameters as those presented in Ref.~\cite{MIST1} (cf. table 1 and section 3 for further explanations). All other parameters have been used with their default values. Table~\ref{tab:MESA} summarizes the value of the relevant parameters used in our study. The choice of atmosphere boundary conditions is primordial because it can drastically modify the position of the tracks in the color axis. \texttt{MESA} offers several methods to compute the surface pressure and temperature. We opted for pre-calculated tables based on \texttt{Atlas} model atmospheres as Ref.~\cite{MIST1} did. Given the masses and the metallicity, effective temperature and surface gravity ranges studied, we found the best fit to be model atmosphere tables for the photosphere.
We also compute stellar tracks with the old stellar code \texttt{JimMacD}~\cite{Jimenez95} in order to show the robustness of the position of the RGB to input physics and how a different modelling of stellar structure doesn't affect our conclusions. While the inital mass value  adopted using the \texttt{MESA} code is  $  M = 0.75 \:M_{\odot}$,  this  exact value is not available for  \texttt{JimMacD}.  Hence, when  comparing the two codes directly,  we also consider  the closest available value in \texttt{JimMacD} which is $  M = 0.80 \:M_{\odot}$, and use initial metallicity $Z = 0.0002$ ([Fe/H]$=-2$). As discussed above, the RGB is very insensitive to the choice of mass.  The detailed comparison of the two codes is presented in  Appendix \ref{sec:stellcode}  where also the corresponding  stellar tracks  are shown. Since the impact of differences between the two codes in our results is negligible,  in what follows the main text only report results for the \texttt{MESA} code.

\begin{table}[h!]
\centering
\begin{threeparttable}
\resizebox{\textwidth}{!}{%
\begin{tabular}{|c|c|}
\hline
\textbf{Parameter}     & \textbf{Value}   \\ \hline
Initial mass $M$  & 0.75, 0.80  \\ \hline
Initial metallicity $Z$  & [0.00005, 0.0004] $\Delta_{Z}=  0.00005$ \\ \hline
Initial Helium mass fraction   & 0.245\tnote{\dag} \\ \hline
Mixing length $\alpha_{\rm MLT}$   &  [1.2, 2.8],  $\Delta_{\alpha}= 0.1$ \\ \hline
Alpha enhancement [$\alpha$/Fe] & 0,  solar scaled \\ \hline
Diffusion and gravitational settling  & Included \\ \hline
Atmosphere boundary conditions & \begin{tabular}{c|c}Model atmosphere tables: & Kurucz \cite{Kurucz1, Kurucz2}\\ Location: & Photosphere \\ Solar abundance: & Asplund et al. \cite{Asplund} \\ Range of $T_{\rm eff}$: & 2500K to 50,000K  \\ Range of $log(g)$: & 0 to 5 \\ Metallicity range [$Z/Z_{\odot}$]: & -7 to 0.5 \end{tabular} \\ \hline
\end{tabular}%
}
\begin{tablenotes}
\item[\dag] Value similar to the initial heliun mass fraction of \texttt{DSED} isochrones.
\end{tablenotes}
\end{threeparttable}
\caption{Values for the stellar parameters used when computing the \texttt{MESA} stellar tracks.}
\label{tab:MESA}
\end{table}

\subsection{Response of the color of the RGB to changes in metallicity and $\alpha_{\rm MLT}$}
\label{sec:dcolordalpha}
The grid of stellar tracks enables us to estimate how changes in key parameters (metallicity and   $\alpha_{\rm MLT}$) affect the color of the RGB. For relatively small changes around fiducial values we can linearize this dependence and report  an estimate of $d {\rm C}/d \alpha_{\rm MLT}$ (where C denotes the color of the RGB at a given magnitude) and  $d {\rm C}/d {\rm Z}$ or $d {\rm C}/d {\rm [Fe/H]}$, obtained as finite differences for few representative magnitudes. These quantites are only indicative, but can help build physical intuition about the effect we want to describe. We find that  these  quantities, as expected,  depend on the magnitude; results are reported in table \ref{tab:derivs}. The color response   to metallicity  ($Z$) is linear to a very good approximation, the response does not depend on the $\Delta Z$ adopted to compute the derivative, but it shows some dependence on the fiducial choice of $\alpha_{\rm MLT}$. We find that  we can approximately  rescale  the derivative to different fiducial values as  $(\alpha_{\rm MLT}/\alpha_{\rm fid})^{3} d {\rm C}/d {\rm Z}|_{\alpha_{\rm fid}}=$ const. if  $\alpha_{\rm fid}$ is around the solar value ($\alpha_{\rm MLT \odot}\sim 2$). 
The relation between $Z$ and [Fe/H] is not linear hence   we report  $\frac{\Delta {\rm C}}{\Delta {\rm[Fe/H]}}= Z \frac{\Delta {\rm C}}{\Delta Z}$ for two represenative metallicities; using the linearized relation  for [Fe/H] is valid only for small shifts.  The $\alpha_{\rm MLT}$ dependence on the other hand is not linear, so the linearized approximation is only valid for small changes  $\Delta_\alpha <0.1$, which is what we adopt here.

\begin{table}[h!]
\centering
\begingroup
\setlength{\tabcolsep}{10pt} 
\renewcommand{\arraystretch}{1.5} 
\begin{tabular}{|c|c|c|c|}
\hline
 &$\Mag=0$ &$\Mag=-2$\\
 \hline
$\frac{\Delta {\rm C}}{\Delta {\rm Z}}|_{Z=0.00015}$ &115 &285\\
$\frac{\Delta {\rm C}}{\Delta {\rm Z}}|_{Z=0.00025}$ &106 &297\\
 $\frac{\Delta {\rm C}}{\Delta {\rm[Fe/H]}}|_{\rm [Fe/H]=-2.0}$ &0.017 &0.043\\
  $\frac{\Delta {\rm C}}{\Delta {\rm[Fe/H]}}|_{\rm [Fe/H]=-1.75}$ &0.027 &0.074\\
 $\frac{\Delta {\rm C}}{\Delta \alpha_{\rm MLT}}(Z=1.5 \times 10^{-4})$ & 0.116 & 0.202\\
 $\frac{\Delta {\rm C}}{\Delta \alpha_{\rm MLT}}(Z=2.5 \times 10^{-4})$ & 0.125 & 0.225\\
\hline
\end{tabular}
\endgroup
\caption{Response  of the RGB color to changes in metallicity and mixing length parameter around a fiducial model for the stellar track of $\alpha=2$.  Here $\Mag$ denotes the magnitude in F606W filter and color, C,  denotes the difference  F606W-F814W. The response to  $\alpha_{\rm MLT}$  show some dependence on the fiducial metallicity so we report several  representative values.}
\label{tab:derivs}
\end{table}

\begin{figure} 
\includegraphics[width=0.5\textwidth]{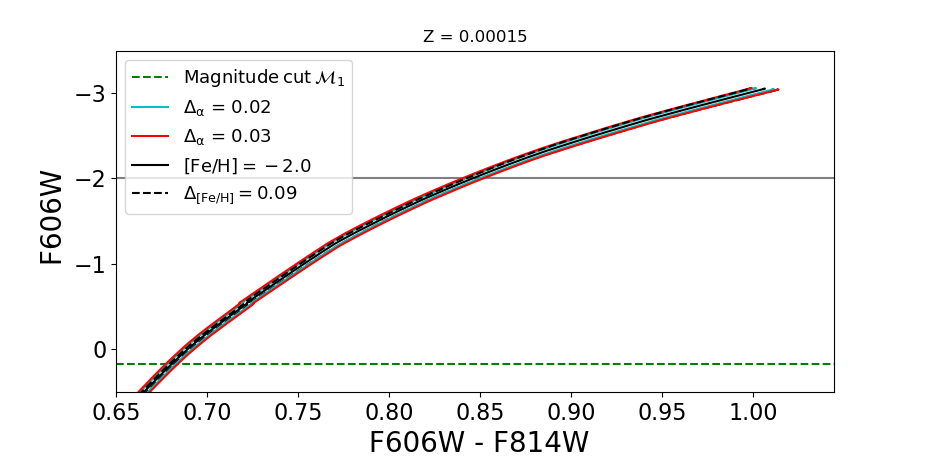} 
\includegraphics[width=0.5\textwidth]{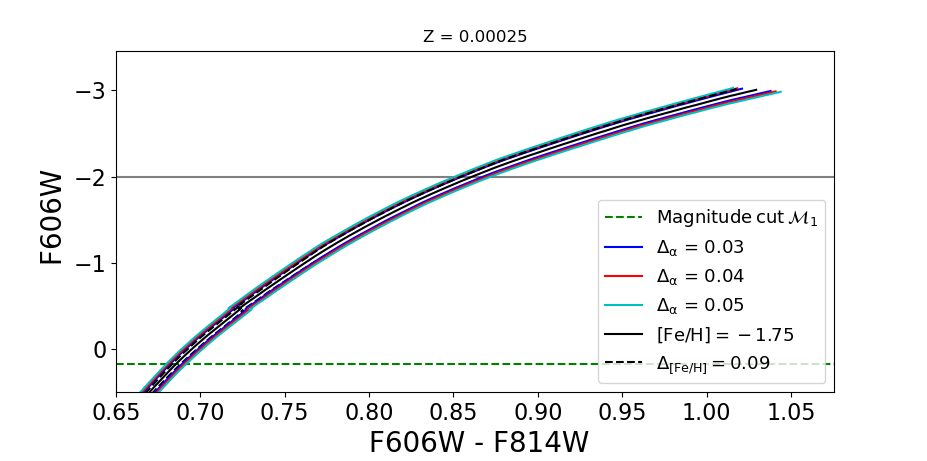} 
\includegraphics[width=0.5\textwidth]{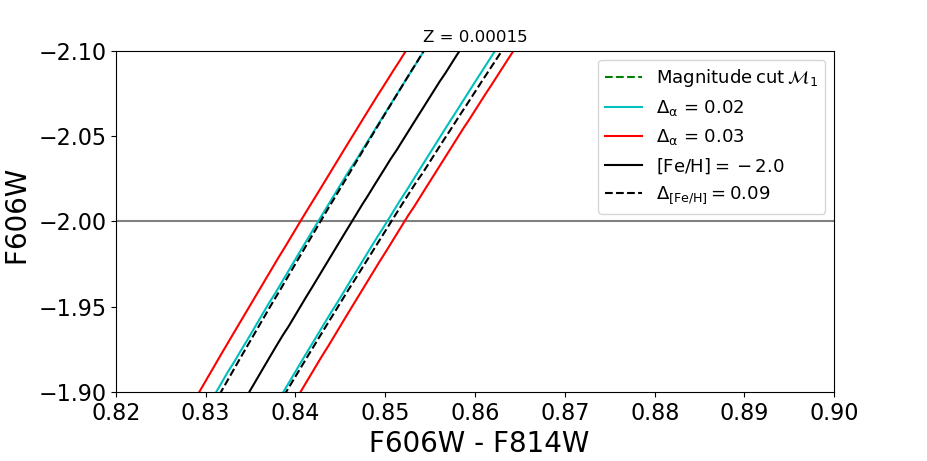} 
\includegraphics[width=0.5\textwidth]{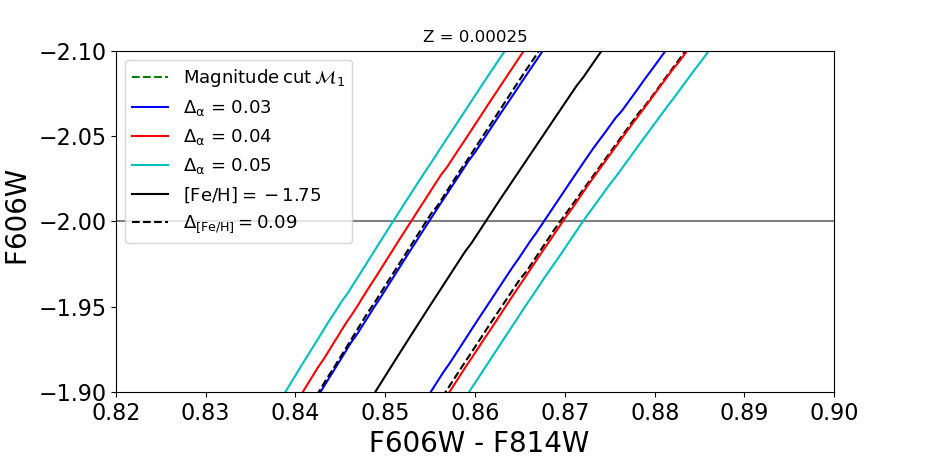} 
\caption{Response of the color of the RGB to a change in metallicity  and a change in $\alpha_{\rm MLT}$ for discrete representative  values. The top panels show the full RGB range while the bottom panels are a zoom in around magnitude $M_1$ (solid  horizontal line). This illustrates that to keep the color of the RGB unchanged for a  small change in  metallicity  $\Delta_{\rm  [Fe/H]}=0.09$, (very close  to the scatter evaluated in sec.~\ref{sec:methods0}, Fig.~\ref{fig:specmetvsmet})  around  the fiducial ($ {\rm [Fe/H]}=-1.75$ and $\alpha_{\rm MLT}=2$), the corresponding change in $\alpha_{\rm MLT}$ is given by $\Delta_\alpha/ \Delta_{\rm  [Fe/H]}\simeq -0.4$. This  is the value we adopt, as it corresponds to that of  the effective metallicity of our GC sample.}
\label{fig:dalphadmet}
\end{figure}

We can then proceed to estimate (approximately, given the linearization assumption implicitly made when computing derivatives by finite differences) what change $\Delta \alpha$ is needed to keep the color of the RGB unchanged under a change in metallicity $\Delta_{\rm [Fe/H]}$. This is visualized in Fig.~\ref{fig:dalphadmet} where one can directly appreciate that a change in metallicity of $\pm \Delta_{\rm [Fe/H]}=0.09$ (which is close to the value of the scatter found in sec.~\ref{sec:methods0})  is compensated by a change $\pm \Delta_{\alpha} \lesssim 0.04$. This upper limit in the required shift in $\alpha_{\rm MLT}$ provides a conservative estimate of the uncertainty in this parameter introduced by its degeneracy with [Fe/H].  This is in broad agreement with the values reported in Tab.~\ref{tab:derivs} if one keeps in mind that the results in the table are approximated because evaluated at a fixed magnitude value implicitly assuming linear dependence of the color on the parameters and  that in practice the overall effect should be seen as a suitably weighted average shift over the magnitude range $\Mag>0$.  In section~\ref{sec:results} we conservatively adopt $d \alpha_{\rm MLT}/d {\rm [Fe/H]}=-0.4$ when converting the measured metallicity scatter  into an estimate fo the scatter in $\alpha_{\rm MLT}$.

\subsubsection{Selecting the RGB}

\begin{figure}
\centering
\includegraphics[width=0.9\textwidth]{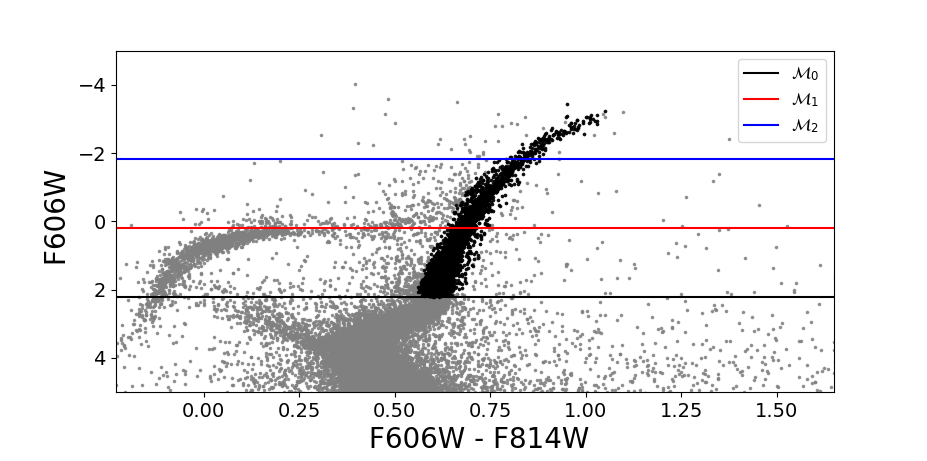} 
\caption{The different cuts in luminosity used to define the spread computed in Table \ref{table:12_dispersion} and Table~\ref{table:alldispersion}. The gray points  represents all the stars in the CMD of  the combined 12 GCs of the $\rm  [Fe/H] < 2$ sample,  the black points show those passing the color  selection of the RGB. }
\label{fig:cuts}
\end{figure}

In order to select only stars in the RGB 
for each cluster we define a band of color around the best fit obtained  in Ref.~\cite{ValcinGC}. We choose a value of $\rm \Delta C$ = 0.06 large enough to include all the stars in the red giant branch and narrow enough to remove most of the stars belonging to the horizontal branch. We also define a magnitude cut, $\Mag_{0}$, corresponding to the start of the RGB. The collection of stars selected for all clusters in the low metallicity sample is shown in Figure \ref{fig:cuts} where the 12 low metallicity  GCs  are plotted on top of each other. Indeed the best fit  $\alpha_{\rm MLT}$ can be biased and the dispersion $\sigma_{\alpha}$ can be  increased by the dispersion in color induced  either by  outliers or by  misclassified stars belonging to the horizontal or asymptotic branch. As the number of stars decreases when we move towards the brightest magnitudes, we define two additional magnitudes cuts to study the dispersion in $\alpha_{\rm MLT}$. The first one $\Mag_{1} = \Mag_{0} - 2.0$ and the second one $\Mag_{2} = \Mag_{1} - 2.0 = \Mag_{0} - 4.0$.  A  value of the scatter in the color of the RGB   changing across different magnitude cuts would indicate  outlier contamination.  Because the correspondence isochrone-stellar tracks is less precise for $\Mag>0$, we adopt results using the $\Mag_1$ cut and report  the $\Mag_0$ cut results only in the Appendices.

\subsubsection{Computing RGB dispersion}
\label{sec:rgbdisp}

We compute the RGB dispersion  for each individual cluster  and the combination of all the GC in each of the three metallicity samples.
If we assume that the photometric dispersion is Gaussian around the isochrone (as argued in Ref.~\cite{ValcinGC}), we can define the dispersion of color by measuring the color distance to the fit: 
\begin{equation}
\rm \sigma_{color} =\: \sqrt{\frac{1}{N}\sum_{i=1}^N (C_{star}^i - C_{fit}^i)^2} 
\end{equation}
 where $\rm C_{star}^i$ and $\rm C_{fit}^i$ respectively correspond to the color of a given star (index $i$) and the color of the track at the same magnitude,  and N is the number of stars in the magnitude interval considered.  
As we compute the dispersion for brighter magnitude cuts ($\Mag_{1}$ and $\Mag_{2}$), the number of stars N decreases and  the distribution becomes sometimes dominated by Poisson noise for individual clusters. Therefore, we set a limit of N = 10 under which the dispersion is not computed.
To estimate the scatter in the mixing length the ideal approach would be to perform a  Bayesian analysis with the mixing length as a free parameter (akin to the approach of Ref.~\cite{ValcinGC}). However  stellar grids are defined only for a specific value of $\alpha_{\rm MLT}$;  recomputing full grids of different mixing lengths and would be very computationally expensive and is beyond the scope of this paper. We proceed instead as follows.  We start by mapping the response of the RGB track to discrete changes in $\alpha_{\rm MLT}$ (see Appendix \ref{sec:stellcode}).  

A (small) shift  in the $\alpha_{\rm MLT}$  value might happen when matching tracks with isochrones. We then  perform a least square fit between stellar tracks for  different values of the mixing length parameter and  the isochrone obtained from  the best fit parameters of  Ref.~\cite{ValcinGC} for each GC. First we select all the evolutionary equivalent points of the isochrone for the magnitude range considered in this work. Then we interpolate all the computed tracks on the same magnitude interval. Finally we compare the tracks and the isochrone at each EEPs magnitude and do a least square fit for $\alpha_{\rm MLT}$. This yields for each cluster a ``best match" $\alpha_{\rm MLT}$ since it is obtained by comparing two different models (track to isochrone). 
For the combination of all the clusters within a metallicity sample, since a best fit isochrone is not available  we do not compare  tracks to the best fit isochrone, instead we  perform a fit  to all the stars in the CMD for the full sample  following  the same fitting procedure as in Ref.~\cite{ValcinGC} but using tracks as the theoretical model instead of isochrones and varying  only $\alpha_{\rm MLT}$. We refer to this  as best fit $\alpha_{\rm MLT}$ since it is obtained via a model to data comparison.

 To compute the scatter  for each cluster, each star in the  RGB  is assigned a value of the mixing length parameter, $\alpha_{\rm MLT}^{\rm i}$, obtained by linear interpolation of the values corresponding to the two closest tracks at the same magnitude, and a corresponding shift $S^{\rm i}$ as the difference between the interpolated  $\alpha_{\rm MLT}^{\rm i}$ and the corresponding best match $\alpha_{\rm MLT}$.

 The dispersion in $\alpha_{\rm MLT}$  is then given by
 \begin{equation}
 \rm \sigma_{\alpha} =\: \sqrt{\frac{1}{N}\sum_{i=1}^N (S^i)^2} \,.
 \end{equation}
 
 For  each sample we take the mean of the scatters of the individual clusters in the sample. The results can be seen in in Table \ref{table:12_dispersion}, but more detailed results on a cluster by cluster basis are reported  in  Appendix \ref{app:allGC}, Table~\ref{table:alldispersion}.

\section{Results}
\label{sec:results}
 The considerations of  Sec.~\ref{sec:methods0}  indicate that the adopted  fiducial value for $\alpha_{\rm MLT}$ adopted by Ref.~\cite{ValcinGC} is not biased. The  findings of  Sec.~\ref{sec:dcolordalpha} and Fig.~\ref{fig:dalphadmet}    yield an estimate for $\Delta_{\alpha}/\Delta_{\rm [Fe/H]}=-0.4$ (the change in $\alpha_{\rm MLT}$ required to compensate a change in  metallicity in order to keep the color of the RBG unchanged) around the fiducial value for $\alpha_{\rm MLT}$ and for the effective value of  [Fe/H] and  magnitude cut.  In Sec.~\ref{sec:methods0} the scatter  between spectroscopic and CMD-estimated  metallicity is estimated to be   $\sigma_{\rm [Fe/H]}=0.093$. 
 If this  is attributed solely to  cluster to cluster (or star to star) changes in  mixing length parameter, we obtain  an upper limit  of $\sigma^{{\rm spec.met}}_{\alpha}=0.04$ (the superscrip stresses that this is computed  resorting to spectroscopic metallicity data). 

Without resorting to external data sets, we can proceed empirically. The spread in color (color scatter) of the RGB for each GC, is  generated by a combination of effects, the dominant one being measurement and photometric errors, as well as  all other stellar parameters variations, which are subdominant. If the color scatter is attributed solely to changes in mixing length parameter it can be used to provide a conservative estimate of star to star variations in $\alpha_{\rm MLT}$. This statement assumes that measurement and photometric errors and variations of stellar parameters are all  random and uncorrelated. In principle, if different sources of scatter are suitably (anti) correlated, this would not necessarily be conservative estimate. We deem this possibility very unlikely.
Combining GC of similar metallicity, the  scatter around the RGB also accounts for possible cluster-to-cluster variations as RGBs of clusters of  similar metallicities are mostly affected by $\alpha_{\rm MLT}$. In this case there is  an additional contribution to the scatter arising from the fact that to convert the observed CMD to the absolute one we have used the best-fit values of  absorption and distance, which may be affected by their own measurement errors. 
       
Results are reported in table~\ref{table:alldispersion}. First note that the scatter in color $\sim 0.02$ is much smaller than the  initial color cut  of $\Delta {\rm C}=0.06$, which clips only the tails of the  color distribution  beyond $3\sigma$s and thus confirming that the initial cut does not affect the estimate of the scatter.
In Appendix ~\ref{app:allGC} we report the results  for the dispersion in color and in mixing length parameter for each cluster and for each of our three sub-samples combined.
 We  note that  all GCs of similar metallicity have a similar value of the dispersion. 
Finally, Table~\ref{table:12_dispersion}  reports the  mixing length best fit and dispersion for each of the metallicity samples. The best fit value for $\alpha_{\rm MLT}$ depends very weekly on metallicity (both the metallicity of the sample and the adopted fiducial metallicity), the scatter does not show any  significant dependence on metallicity. 
The scatter for each metallicity sample (which include $\gtrsim 10$ clusters) is comparable with the individual cluster scatter indicating that there is not additional cluster-to-cluster variation.  We adopt a, suitably weighted combined scatter across the three metallicity samples of $\sigma^{\rm CMD}_{\alpha}=0.15$ as a conservative estimate of the $\alpha_{\rm MLT}$ scatter estimated from the CMD of the clusters in the sample. Recall that we have attributed the full  color scatter of the RGB to $\alpha_{\rm MLT}$,  when the color scatter include contributions from measurement errors, photometric errors as well as variations of all other model parameters.

A suite of tests ensuring the robustness of these results to several commonly adopted assumptions  is presented in  the appendices. In particular Appendix \ref{sec:metmass} discusses the impact of  chosen fiducial values of mass and metallicity;  Appendix~\ref{sec:stellcode} compares different stellar codes and  Appendix ~\ref{sec:robustnesstests}  tests the effects of microphysics modelling. 
In summary, there is no indication that stars in old GCs  have values for the  depth of the convection envelope  different from those obtained for the Sun. For the 38 GC considered here we find that the  preferred value for the mixing length parameter is $\alpha_{\rm MLT}=1.90 \pm 0.04\, ({\rm or} \pm 0.15)$ with (without) resorting to spectroscopic metallicity determinations,  where the reported error is  estimated conservatively. 

\begin{table}[]
\centering
\resizebox{0.65\textwidth}{!}{%
\begin{tabular}{|c|c|c|c|c|}
\hline
\multicolumn{5}{|c|}{\textbf{$\rm [Fe/H] < -2.0, 12\: clusters$}} \\ \hline
Metallicity  & $\alpha (\Mag_1) $ best fit   &$\alpha (\Mag_1) $ best match   & $\sigma_{\alpha} (\Mag_{1})$ & $\sigma_{\alpha} (\Mag_{2})$     \\ \hline
$Z = 0.00005$ & 1.9 & 1.89 & 0.17 & 0.08 \\ \hline
$Z = 0.00010$ & 1.9 & 1.95 & 0.17 & 0.1 \\ \hline
$Z = 0.00015$ **& 2.0 & 1.98 & 0.17 & 0.08 \\ \hline
$Z = 0.00020$ & 2.0 & 2.03 &  0.17 & 0.08 \\ \hline
\hline
\multicolumn{5}{|c|}{\textbf{$\rm -2.0 < [Fe/H] < -1.75, 11\: clusters$}} \\ \hline
Metallicity  & $\alpha (\Mag_1) $ best fit   &$\alpha (\Mag_1)$ best match  &  $\sigma_{\alpha} (\Mag_{1})$ & $\sigma_{\alpha} (\Mag_{2})$     \\ \hline
$Z = 0.00015$ & 1.8 & 1.81 &  0.14 & 0.07 \\ \hline
$Z = 0.00020$** & 1.9 & 1.85 & 0.14 & 0.08 \\ \hline
$Z = 0.00025$** & 1.9 & 1.89 & 0.14 & 0.07 \\ \hline
$Z = 0.00030$ & 1.9 & 1.93  & 0.13 & 0.08  \\ \hline
\hline
\multicolumn{5}{|c|}{\textbf{$\rm -1.75 < [Fe/H] < -1.50, 15\: clusters$}} \\ \hline
Metallicity  & $\alpha (\Mag_1) $ best fit   &$\alpha (\Mag_1)$ best match & $\sigma_{\alpha} (\Mag_{1})$ & $\sigma_{\alpha} (\Mag_{2})$     \\ \hline
$Z = 0.00025$ & 1.8 & 1.81 &  0.14 & 0.07 \\ \hline
$Z = 0.00030$** & 1.8 & 1.85 & 0.14 & 0.08  \\ \hline
$Z = 0.00035$ & 1.9 & 1.89 &  0.13 & 0.07 \\ \hline
$Z = 0.00040$ & 1.9 & 1.91 & 0.13 & 0.07  \\ \hline
\end{tabular}%
}
\caption{Summary of the mixing length best fit and best match for the $\Mag_1$ cut and dispersion for each of the metallicity samples (full information  for individual clusters is available in appendix \ref{app:allGC}). Each sample spans a range in metallicities, and we report results for several  fiducial metallicity values covering the range. The values closer to the effective [Fe/H]  metallicity of the sample are flagged by the asterisks. The best fit value for $\alpha_{\rm MLT}$ depends very weekly on metallicity (both the metallicity of the sample and the adopted fiducial metallicity), the scatter does not show any  significant dependence on metallicity. }
\label{table:12_dispersion}
\end{table}
\section{Conclusions and implications for the age of the Universe}
\label{sec:summary}

 An estimate of the age of the Universe from the age of the oldest globular clusters was presented in Ref.~\cite{ValcinGC}: the age of the oldest clusters  being $t_{\rm GC}=13.32\pm0.10({\rm stat.})\pm 0.5({\rm sys.})$ Gyr, and the inferred age of the Universe $t_{\rm U}=13.5^{+0.16}_{-0.14}({\rm stat.}) \pm 0.5({\rm sys.})$ Gyr. The dominant contribution to the error on this quantity is due to  the systematic uncertainty  in the depth of the convention envelope (the mixing length parameter) accounting for 0.3  (i.e. 60\%) of the 0.5 systematic error budget.   Ref.~\cite{ValcinGC} adopted a range in $\alpha_{\rm MLT}$  corresponding to the full range  considered in  Ref.~\cite{OMalley} which in the convention of this paper corresponds to  $2\Delta_{\alpha}=0.7$.
  
 Here we have studied the dependence of the morphology of the RGB in the GCs CMD on changes in $\alpha_{\rm MLT}$, in order to provide a  more realistic estimate of the uncertainty on this parameter. We have shown that the range used in Ref.~\cite{ValcinGC} include values that do not fit the observed properties of the GCs in our sample. After studying the degeneracy between $\alpha_{\rm MLT}$ and metallicity, we have estimated an upper limit for the uncertainty of  $\alpha_{\rm MLT}$ for our sample:   $\sigma^{\rm spec.met}_{\alpha}=0.04$ or $\sigma^{\rm CMD}_{\alpha}=0.15$   (depending  whether using external spectroscopic metallicity determinations  or not).  It is interesting to note that recently, Ref.~\citep{Tognelli} performed a Bayesian calibration of the mixing length parameter $\alpha$ using mock and real data of the Hyades open cluster and found an average value $\langle\alpha\rangle \:= 2.01 \pm 0.05$. This result is in good agreement with our findings.
 
With this reduction of the dominant systematic contribution  to the age determination (from 0.3 to 0.13 or 0.034 Gyr), the  mixing length parameter  cease to be  the dominant contribution to the uncertainty; now the leading systematic uncertainties are due to nuclear reaction rates and opacities.

Thanks to the reduction in the systematic error budget achieved in this work, we conclude\footnote{To propagate exactly an error in $\alpha_{\rm MLT}$ into an error in age one would have to reproduce the full calculations of Ref.~\cite{OMalley}, which goes well beyond the scope of this paper.   Ref.~\cite{OMalley} shows that  $\Delta_{\alpha}=\pm 0.35$ yields a $\Delta_{\rm age}=\pm 0.3$. Here we simply rescale the  age uncertainty according to our newly determined $\alpha_{\rm MLT}$ uncertainty.} that the age of the oldest globular clusters is   $t_{\rm GC}=13.32\pm0.10({\rm stat.})\pm 0.23(0.33)({\rm sys.})$ Gyr, which corresponds to an age of the Universe of  $t_{\rm U}=13.5^{+0.16}_{-0.14}({\rm stat.})\pm 0.23(0.33)({\rm sys.})$ Gyr., an uncertainty of 0.27(0.36) Gyr if statistical and systematic errors are added in quadrature. 
This determination of the age of the Universe is cosmological-model agnostic, in the sense that it does not depend in any significant way on the cosmological model adopted, and is in good agreement with the cosmological model-dependent determination of $t_{\rm U}=13.8\pm 0.02$ Gyr inferred from the Planck mission from observations of the Universe at $z\sim1100$,  assuming the standard $\Lambda$CDM model. The implications for cosmology of this consideration are  explored in a companion paper \cite{TrianglesPaper}.

\acknowledgments

We thank the referee for a timely, useful, constructive and insightful feedback.
We  also thank the stellar modelers for making their stellar models publicly available. This work is supported by  MINECO grant PGC2018-098866-B-I00 FEDER, UE.    LV acknowledges support by European Union's Horizon 2020 research and innovation program ERC (BePreSySe, grant agreement 725327). JLB is supported by the Allan C. and Dorothy H. Davis Fellowship. 
The work of BDW is supported by the Labex ILP (reference ANR-10-LABX-63) part of the Idex SUPER,  received financial state aid managed by the Agence Nationale de la Recherche, as part of the programme Investissements d'avenir under the reference ANR-11-IDEX-0004-02; and by the ANR BIG4 project, grant ANR-16-CE23-0002 of the French Agence Nationale de la Recherche. 
The Center for Computational Astrophysics is supported by the Simons Foundation. 

\appendix

\begin{figure}[!h]
\centering
\includegraphics[width=0.9\textwidth]{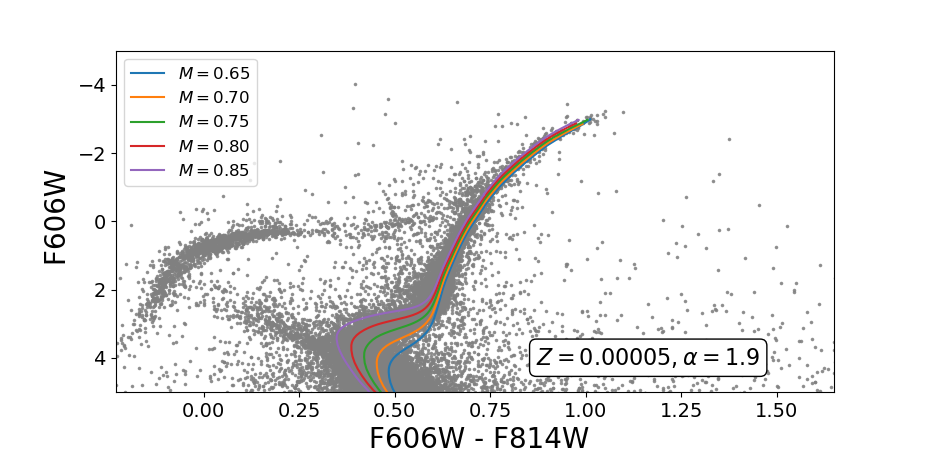} 
\caption{Effect of varying mass for a star with a fixed metallicity and mixing length.  The grey points represent the  CMD of all the clusters in the low metallicity sample. The lines are stellar tracks for an age of 13.32 Gyr  corresponding to different initial masses. If the tracks are interpreted as isochrones such a spread in mass would correspond to a (widely unrealistic) range in age from 11 to 30 Gyr.  Note that the RGB morphology is very insensitive to mass (and age). The $\alpha_{\rm MLT}$ value adopted here is 1.9 as it is the closest in our grid to the \texttt{DSED} code solar value.}
\label{fig:mass}
\end{figure}

\begin{figure}
\centering 
\includegraphics[width=\textwidth]{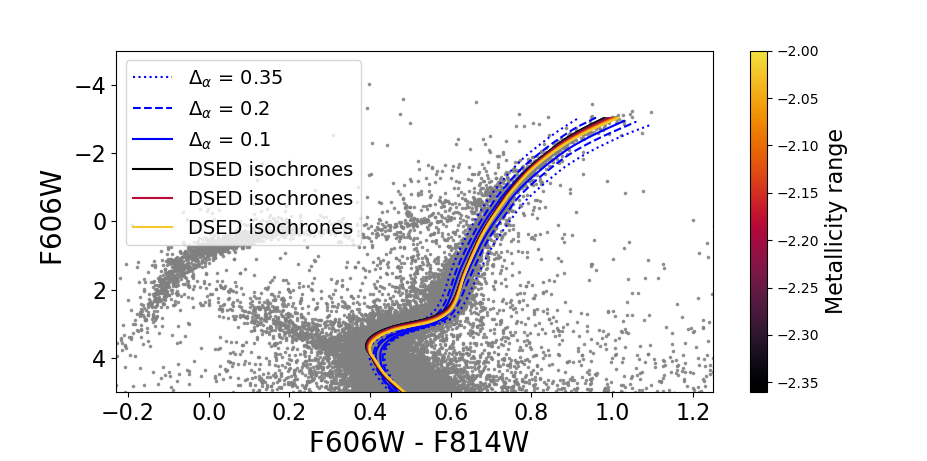} 
\includegraphics[width=\textwidth]{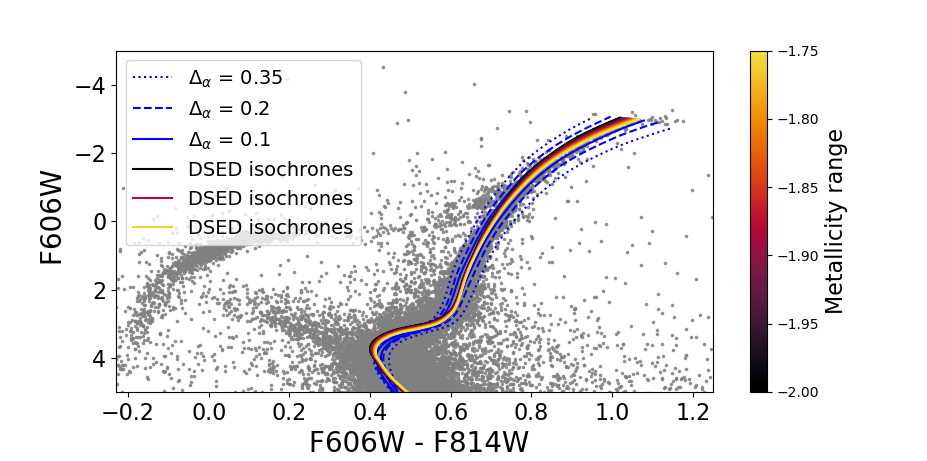}
\caption{Effect of varying metallicity for a track with a fixed mass and mixing length, compared to a change in  mixing length for fixed mass. Top panel: Grey points: the combined CMD of the sample of  12 clusters with metallicity {Fe/H}$<-2$. Solid lines tracks color-coded by metallicity for a range spanning the low metallicity sample.  Blue lines: effects of changes in $\alpha_{\rm MLT}$. Bottom pane: as for top panel but for the intermediate metallicity sample.}
\label{fig:metal}
\end{figure}

\section{Impact of mass, metallicity and  $\alpha_{\rm MLT}$ on the RGB}
\label{sec:metmass}

As mentioned in section \ref{tracks}, mass and metallicity are key   parameters  for our purpose of  constraining the mixing length parameter. It is possible to compare the effect of mass (for a track) to the effect of age (for an isochrone). The   initial mass of the star influences the time spent on the main sequence but affects very little the red giant branch.  In figure~\ref{fig:mass} we explore a range of masses  from $M = 0.65$ to $0.85$ $ M_{\odot}$ with $\Delta_{M}$ = 0.05. 

Metallicity affects the color of the RGB: an increase in metallicity will result in a tilt of the RGB towards higher ${\rm C}$. In Figure \ref{fig:metal} we show the effect of varying metallicity for a track with a fixed mass and mixing length, compared to a change in  mixing length for fixed mass. This figure illustrates that for a  metallicity interval comparable with that of each of our samples, the mixing length is the determining parameter in the color of the RGB.

\begin{figure}[h!]
\centering
\includegraphics[width=0.9\textwidth]{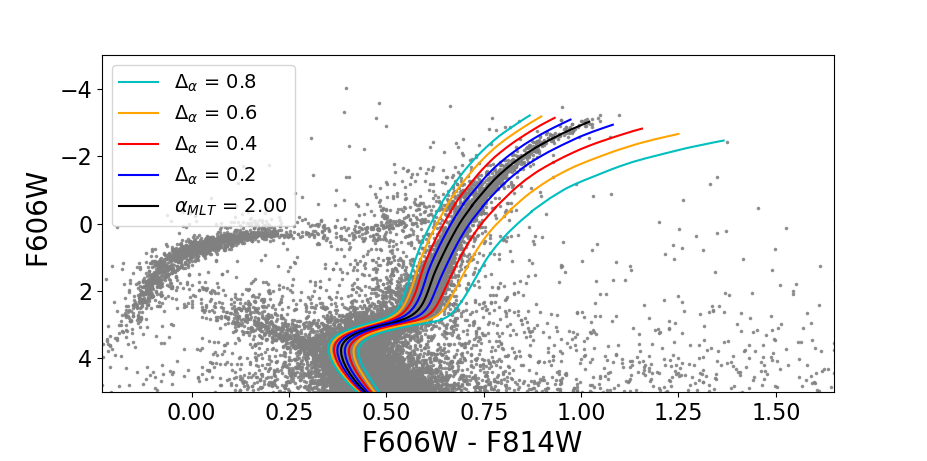} 
\includegraphics[width=0.9\textwidth]{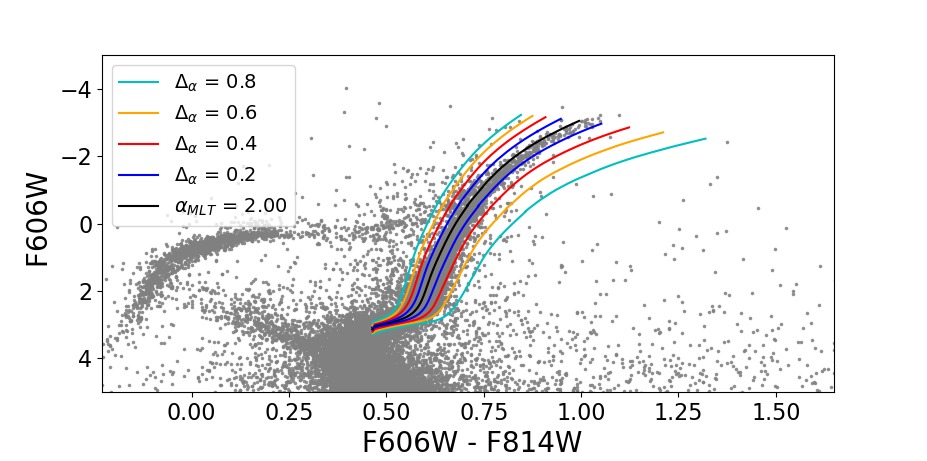} 
\caption{Stellar tracks are shown for a star with initial mass 0.80 $M_{\odot}$ and $Z = 0.0002 $ for the MESA code and several values of the mixing length around $\alpha=2$ .  The spread of the RGB  roughly corresponds to  $\Delta_\alpha \sim  0.1$. Top panel: Using the Casagrande \& VandenBerg \cite{Casagrande} bolometric correction, Bottom panel: Using bolometric correction computed from the Castelli \& Kurucz 2003 \cite{Castelli}  atmospheric spectra. on the relevant part of the RGB the two plots are virtually indistinguishable.}
\label{fig:fit}
\end{figure}

The value of  the $\alpha_{\rm MLT}$ parameter is stellar code dependent. The  conversion between different code-conventions is just a shift,  while the relevant quantity  for our argument here is the interval or range adopted which is convention-independent. 
In Ref.~\cite{ValcinGC} we used the isochrones from the \texttt{DSED} model for a solar value for $\alpha_{\rm MLT}$  and here stellar tracks from an independent code. An even slightly incorrect conversion would result into adopting an incorrect fiducial  $\alpha_{\rm MLT}$  value and possibly an over-estimate of the  color scatter.
To map the response of the RGB to changes in mixing length parameter values, we compute tracks for discrete values of $\alpha_{\rm MLT}$ (from 1.2 to 2.8 in steps of 0.1) as illustrated in Figure \ref{fig:fit} where the tracks are centered around $\alpha_{\rm MLT}= 2$.

Bolometric corrections are used to transform from observed colors to theoretical effective temperature and vice-versa. While a set of bolometric correction tables is provided for the pair \texttt{MIST} - \texttt{MESA}, there is none offered for the couple \texttt{DSEP} - \texttt{DSED}. The authors thought of using the \texttt{MIST} tables but these are only available for solar-scaled calibration and we also wanted to test tables with alpha enhancement. Even if, in the end, the solar-scaled calibration gave us a better fit as opposed to the other choice of [$\alpha/\rm Fe] = 0.4$ (probably because most of the clusters selected have [$\alpha/\rm Fe]$ best fits between 0.1 and 0.2), we continued to work with the same tables. These corrections, if not accurate enough, can lead to an additional systematic uncertainty. To quantify this effect, we use two different sets of bolometric corrections \cite{Casagrande,Castelli}. As it is shown in Fig.~\ref{fig:fit} both lead to the same transformation of colors to effective temperature. We therefore do not propagate any additional uncertainty due to bolometric corrections.

\begin{figure}
\centering
\includegraphics[width=\textwidth]{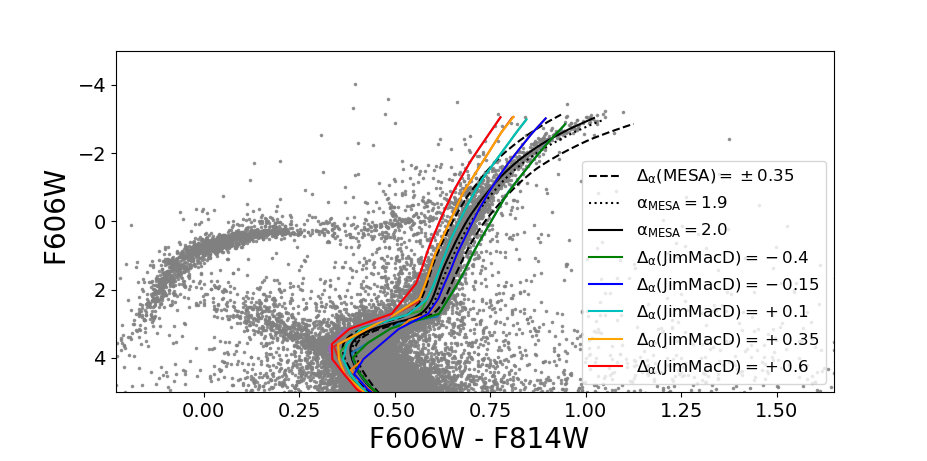}  
\caption{Response of the RGB to changes in $\alpha_{\rm MLT}$ across different codes. As before, the gray points correspond to all the stars in the CMD of the 12 clusters in the low metallicity sample.  The black  solid line  corresponds to \texttt{MESA} track for $\alpha_{\rm MLT}=2$, $M=0.8M_{\odot}$, $Z=0.00002$, the dotted line to $\alpha_{\rm MLT}=1.9$ for reference, and the dashed lines illustrate a change $\Delta_\alpha=\pm0.35$ around the solar value. The solid lines in colors correspond to \texttt{JimMcD} tracks for the same mass and metallicities and for shifts from the solar  value as indicated in the legend. The poor fit at the
very tip of the RGB is due to the lack of cool opacities in the code unlike the modern \texttt{MESA} stellar code.}
\label{fig:fitMacD}
\end{figure}

\section{Effect of opacities and nuclear reaction rates}
\label{sec:stellcode}
Uncertainties in opacities and nuclear reaction rates  also  contribute to the systematic error budget on age determinations, but  are expected to be sub-dominant compared to the effect of the mixing length parameter. Here we illustrate this by resorting to available outputs of a different stellar code  where these quantities are different from those assumed in \texttt{MESA}. In this case we chose the old version (1996) of the \texttt{JimMacD} code and stellar tracks \cite{Jimenez95}, which uses very different stellar opacities, boundary conditions and mixing-length formulation.    \texttt{JimMacD} tracks are available for initial mass 0.8 $M_{\odot}$ (hence we compare  directly with \texttt{MESA} for this value of the mass) and for a coarse grid of $\alpha_{\rm MLT}$ values (recall that the solar value for $\alpha_{\rm MLT}$ for  \texttt{JimMacD} is 1.4, which is however not available on the provided grid).  This is shown in Fig.~\ref{fig:fitMacD}. Note  that the  \texttt{JimMacD} code fails to fit the tip of the RGB because of the lack of cold opacities, not available at that time.  Where this is not a relevant effect, tracks for solar values of the mixing langth parameters agree well, and  the  response of the RGB color to changes in $\alpha_{\rm MLT}$ is very similar across the two codes: we find that for $\Mag = 0$ and for the same mass  0.8 $M_{\odot}$, metallicity  and $\alpha_{\rm MLT}$ step sampled by  \texttt{JimMacD} tracks, $\Delta {\rm C}/\Delta \alpha_{\rm MLT}=-0.12$ and $-0.13$ for \texttt{MESA} and \texttt{JimMacD} respectively. 
We therefore conclude that the differences do not  significantly bias the RGB color or consequently the recovery of an $\alpha_{\rm MLT}$ value consistent with solar, further supporting the robustness of the results reported in the main text. 

\section{Parameter constrains for  all  GCs}
\label{app:allGC}

We double check that the $\alpha_{\rm MLT}$  that best fits the RGB  is  consistent with the adopted fiducial value of Ref.~\cite{ValcinGC} and consequently the fiducial adopted here too.   An incorrect fiducial  $\alpha_{\rm MLT}$  value  would yield to biases and possibly an over-estimate of the  color scatter. Table \ref{table:alldispersion} reports  for each cluster the color scatter as a function of magnitude cut,  the best match $\alpha_{\rm MLT}$, the scatter in this quantity as a function of magnitude cut and fiducial assumed metallicity. The table also report the best fit $\alpha_{\rm MLT}$ for the combination of the GC in each sample and the recomputed scatter with respect to this quantity rather than the best match $\alpha_{\rm MLT}$. The differences, however are unimportant.

The tables presented here  compement Table~\ref{table:12_dispersion} in the main text. The  best fit $\alpha_{\rm MLT}$ is  well consistent with the solar value adopted by \texttt{DSED} ($\alpha_{\rm MLT}=1.938$), validating  this choice  for  the fiducial value assumed with no evidence for any bias.

\begin{table}[]
\huge
\centering
\resizebox{\textwidth}{!}{%
\begin{tabular}{ccccccccccccccccccc}
                                   &                             &                             & \multicolumn{4}{c}{}                                                                                          & \multicolumn{4}{c}{}                                                                                          & \multicolumn{4}{c}{}                                                                                          &                           &                           &                           &                           \\ \hline
\multicolumn{19}{|c|}{Sample 1, 12 clusters  with  {[}Fe/H{]} \textless 2.0}                                                                                                                                                                                                                                                                                                                                                                                                                                                                                   \\ \hline
                                   &                             &                             &                           &                           &                           &                           &                           &                           &                           &                           &                           &                           &                           &                           &                           &                           &                           &                           \\ \hline
&       & \multicolumn{1}{c|}{}       & \multicolumn{4}{c|}{$Z = 0.00005$}                                                                              & \multicolumn{4}{c|}{$Z = 0.00010$}                                                                              & \multicolumn{4}{c|}{$Z = 0.00015$}                                                                              & \multicolumn{4}{c|}{$Z = 0.00020$}                                                                              \\ \hline
\multicolumn{1}{|c|}{GC name }    & \multicolumn{1}{c|}{$\sigma_{color} (\Mag_{0})$}  & \multicolumn{1}{c|}{$\sigma_{color} (\Mag_{1})$}&  \multicolumn{1}{c|}{$\alpha$ best match} & \multicolumn{1}{c|}{$\sigma_{\alpha} (\Mag_{0})$} & \multicolumn{1}{c|}{$\sigma_{\alpha} (\Mag_{1})$} & \multicolumn{1}{c|}{$\sigma_{\alpha} (\Mag_{2})$}  & \multicolumn{1}{c|}{$\alpha$ best match} & \multicolumn{1}{c|}{$\sigma_{\alpha} (\Mag_{0})$} & \multicolumn{1}{c|}{$\sigma_{\alpha} (\Mag_{1})$} & \multicolumn{1}{c|}{$\sigma_{\alpha} (\Mag_{2})$}  & \multicolumn{1}{c|}{$\alpha$ best match} & \multicolumn{1}{c|}{$\sigma_{\alpha} (\Mag_{0})$} & \multicolumn{1}{c|}{$\sigma_{\alpha} (\Mag_{1})$} & \multicolumn{1}{c|}{$\sigma_{\alpha} (\Mag_{2})$}   & \multicolumn{1}{c|}{$\alpha$ best match} & \multicolumn{1}{c|}{$\sigma_{\alpha} (\Mag_{0})$} & \multicolumn{1}{c|}{$\sigma_{\alpha} (\Mag_{1})$} & \multicolumn{1}{c|}{$\sigma_{\alpha} (\Mag_{2})$}     \\ \hline
\multicolumn{1}{|c|}{NGC 2298}     & \multicolumn{1}{c|}{0.0231} & \multicolumn{1}{c|}{0.0236} & \multicolumn{1}{c|}{1.8}  & \multicolumn{1}{c|}{0.22} & \multicolumn{1}{c|}{0.19} & \multicolumn{1}{c|}{N/A}  & \multicolumn{1}{c|}{1.9}  & \multicolumn{1}{c|}{0.22} & \multicolumn{1}{c|}{0.16} & \multicolumn{1}{c|}{N/A}  & \multicolumn{1}{c|}{1.9}  & \multicolumn{1}{c|}{0.22} & \multicolumn{1}{c|}{0.17} & \multicolumn{1}{c|}{N/A}  & \multicolumn{1}{c|}{2}    & \multicolumn{1}{c|}{0.21} & \multicolumn{1}{c|}{0.16} & \multicolumn{1}{c|}{N/A}  \\ \hline
\multicolumn{1}{|c|}{NGC 4590}     & \multicolumn{1}{c|}{0.0198} & \multicolumn{1}{c|}{0.0158} & \multicolumn{1}{c|}{1.9}  & \multicolumn{1}{c|}{0.21} & \multicolumn{1}{c|}{0.14} & \multicolumn{1}{c|}{N/A}  & \multicolumn{1}{c|}{2}    & \multicolumn{1}{c|}{0.19} & \multicolumn{1}{c|}{0.13} & \multicolumn{1}{c|}{N/A}  & \multicolumn{1}{c|}{2}    & \multicolumn{1}{c|}{0.2}  & \multicolumn{1}{c|}{0.13} & \multicolumn{1}{c|}{N/A}  & \multicolumn{1}{c|}{2.1}  & \multicolumn{1}{c|}{0.18} & \multicolumn{1}{c|}{0.14} & \multicolumn{1}{c|}{N/A}  \\ \hline
\multicolumn{1}{|c|}{NGC 4833}     & \multicolumn{1}{c|}{0.018}  & \multicolumn{1}{c|}{0.0196} & \multicolumn{1}{c|}{1.8}  & \multicolumn{1}{c|}{0.16} & \multicolumn{1}{c|}{0.15} & \multicolumn{1}{c|}{0.09} & \multicolumn{1}{c|}{1.9}  & \multicolumn{1}{c|}{0.15} & \multicolumn{1}{c|}{0.14} & \multicolumn{1}{c|}{0.08} & \multicolumn{1}{c|}{1.9}  & \multicolumn{1}{c|}{0.16} & \multicolumn{1}{c|}{0.14} & \multicolumn{1}{c|}{0.09} & \multicolumn{1}{c|}{2}    & \multicolumn{1}{c|}{0.16} & \multicolumn{1}{c|}{0.14} & \multicolumn{1}{c|}{0.08} \\ \hline
\multicolumn{1}{|c|}{NGC 5053}     & \multicolumn{1}{c|}{0.0155} & \multicolumn{1}{c|}{0.0195} & \multicolumn{1}{c|}{1.9}  & \multicolumn{1}{c|}{0.15} & \multicolumn{1}{c|}{0.14} & \multicolumn{1}{c|}{N/A}  & \multicolumn{1}{c|}{2}    & \multicolumn{1}{c|}{0.18} & \multicolumn{1}{c|}{0.16} & \multicolumn{1}{c|}{N/A}  & \multicolumn{1}{c|}{2}    & \multicolumn{1}{c|}{0.16} & \multicolumn{1}{c|}{0.15} & \multicolumn{1}{c|}{N/A}  & \multicolumn{1}{c|}{2.1}  & \multicolumn{1}{c|}{0.21} & \multicolumn{1}{c|}{0.18} & \multicolumn{1}{c|}{N/A}  \\ \hline
\multicolumn{1}{|c|}{NGC 6341}     & \multicolumn{1}{c|}{0.0172} & \multicolumn{1}{c|}{0.0175} & \multicolumn{1}{c|}{1.9}  & \multicolumn{1}{c|}{0.17} & \multicolumn{1}{c|}{0.15} & \multicolumn{1}{c|}{0.05} & \multicolumn{1}{c|}{2}    & \multicolumn{1}{c|}{0.16} & \multicolumn{1}{c|}{0.16} & \multicolumn{1}{c|}{0.08} & \multicolumn{1}{c|}{2}    & \multicolumn{1}{c|}{0.16} & \multicolumn{1}{c|}{0.15} & \multicolumn{1}{c|}{0.04} & \multicolumn{1}{c|}{2}    & \multicolumn{1}{c|}{0.16} & \multicolumn{1}{c|}{0.15} & \multicolumn{1}{c|}{0.04} \\ \hline
\multicolumn{1}{|c|}{NGC 6397}     & \multicolumn{1}{c|}{0.0254} & \multicolumn{1}{c|}{0.0277} & \multicolumn{1}{c|}{1.9}  & \multicolumn{1}{c|}{0.21} & \multicolumn{1}{c|}{0.2}  & \multicolumn{1}{c|}{N/A}  & \multicolumn{1}{c|}{1.9}  & \multicolumn{1}{c|}{0.22} & \multicolumn{1}{c|}{0.22} & \multicolumn{1}{c|}{N/A}  & \multicolumn{1}{c|}{1.9}  & \multicolumn{1}{c|}{0.24} & \multicolumn{1}{c|}{0.24} & \multicolumn{1}{c|}{N/A}  & \multicolumn{1}{c|}{2}    & \multicolumn{1}{c|}{0.22} & \multicolumn{1}{c|}{0.2}  & \multicolumn{1}{c|}{N/A}  \\ \hline
\multicolumn{1}{|c|}{NGC 6426}     & \multicolumn{1}{c|}{0.0239} & \multicolumn{1}{c|}{0.0287} & \multicolumn{1}{c|}{1.9}  & \multicolumn{1}{c|}{0.23} & \multicolumn{1}{c|}{0.24} & \multicolumn{1}{c|}{N/A}  & \multicolumn{1}{c|}{2}    & \multicolumn{1}{c|}{0.22} & \multicolumn{1}{c|}{0.21} & \multicolumn{1}{c|}{N/A}  & \multicolumn{1}{c|}{2}    & \multicolumn{1}{c|}{0.22} & \multicolumn{1}{c|}{0.22} & \multicolumn{1}{c|}{N/A}  & \multicolumn{1}{c|}{2.1}  & \multicolumn{1}{c|}{0.21} & \multicolumn{1}{c|}{0.2}  & \multicolumn{1}{c|}{N/A}  \\ \hline
\multicolumn{1}{|c|}{NGC 6779}     & \multicolumn{1}{c|}{0.0241} & \multicolumn{1}{c|}{0.0258} & \multicolumn{1}{c|}{1.9}  & \multicolumn{1}{c|}{0.21} & \multicolumn{1}{c|}{0.18} & \multicolumn{1}{c|}{0.09} & \multicolumn{1}{c|}{1.9}  & \multicolumn{1}{c|}{0.22} & \multicolumn{1}{c|}{0.19} & \multicolumn{1}{c|}{0.11} & \multicolumn{1}{c|}{2}    & \multicolumn{1}{c|}{0.2}  & \multicolumn{1}{c|}{0.17} & \multicolumn{1}{c|}{0.09} & \multicolumn{1}{c|}{2}    & \multicolumn{1}{c|}{0.21} & \multicolumn{1}{c|}{0.18} & \multicolumn{1}{c|}{0.1}  \\ \hline
\multicolumn{1}{|c|}{NGC 7078}     & \multicolumn{1}{c|}{0.0219} & \multicolumn{1}{c|}{0.0204} & \multicolumn{1}{c|}{2}    & \multicolumn{1}{c|}{0.19} & \multicolumn{1}{c|}{0.15} & \multicolumn{1}{c|}{0.08} & \multicolumn{1}{c|}{2}    & \multicolumn{1}{c|}{0.2}  & \multicolumn{1}{c|}{0.16} & \multicolumn{1}{c|}{0.08} & \multicolumn{1}{c|}{2.1}  & \multicolumn{1}{c|}{0.18} & \multicolumn{1}{c|}{0.15} & \multicolumn{1}{c|}{0.08} & \multicolumn{1}{c|}{2.1}  & \multicolumn{1}{c|}{0.18} & \multicolumn{1}{c|}{0.15} & \multicolumn{1}{c|}{0.07} \\ \hline
\multicolumn{1}{|c|}{NGC 7099}     & \multicolumn{1}{c|}{0.0268} & \multicolumn{1}{c|}{0.0211} & \multicolumn{1}{c|}{1.9}  & \multicolumn{1}{c|}{0.25} & \multicolumn{1}{c|}{0.16} & \multicolumn{1}{c|}{0.11} & \multicolumn{1}{c|}{1.9}  & \multicolumn{1}{c|}{0.27} & \multicolumn{1}{c|}{0.18} & \multicolumn{1}{c|}{0.14} & \multicolumn{1}{c|}{2}    & \multicolumn{1}{c|}{0.22} & \multicolumn{1}{c|}{0.14} & \multicolumn{1}{c|}{0.09} & \multicolumn{1}{c|}{2}    & \multicolumn{1}{c|}{0.23} & \multicolumn{1}{c|}{0.14} & \multicolumn{1}{c|}{0.12} \\ \hline
\multicolumn{1}{|c|}{Palomar 15}   & \multicolumn{1}{c|}{0.0285} & \multicolumn{1}{c|}{0.0284} & \multicolumn{1}{c|}{1.9}  & \multicolumn{1}{c|}{0.27} & \multicolumn{1}{c|}{0.21} & \multicolumn{1}{c|}{N/A}  & \multicolumn{1}{c|}{2}    & \multicolumn{1}{c|}{0.24} & \multicolumn{1}{c|}{0.17} & \multicolumn{1}{c|}{N/A}  & \multicolumn{1}{c|}{2}    & \multicolumn{1}{c|}{0.25} & \multicolumn{1}{c|}{0.18} & \multicolumn{1}{c|}{N/A}  & \multicolumn{1}{c|}{2}    & \multicolumn{1}{c|}{0.26} & \multicolumn{1}{c|}{0.2}  & \multicolumn{1}{c|}{N/A}  \\ \hline
\multicolumn{1}{|c|}{Terzan 8}     & \multicolumn{1}{c|}{0.0169} & \multicolumn{1}{c|}{0.0189} & \multicolumn{1}{c|}{1.9}  & \multicolumn{1}{c|}{0.17} & \multicolumn{1}{c|}{0.14} & \multicolumn{1}{c|}{N/A}  & \multicolumn{1}{c|}{1.9}  & \multicolumn{1}{c|}{0.17} & \multicolumn{1}{c|}{0.14} & \multicolumn{1}{c|}{N/A}  & \multicolumn{1}{c|}{2}    & \multicolumn{1}{c|}{0.18} & \multicolumn{1}{c|}{0.15} & \multicolumn{1}{c|}{N/A}  & \multicolumn{1}{c|}{2}    & \multicolumn{1}{c|}{0.17} & \multicolumn{1}{c|}{0.14} & \multicolumn{1}{c|}{N/A}  \\ \hline
\multicolumn{1}{|c|}{All 12 GCs}   & \multicolumn{1}{c|}{0.0226} & \multicolumn{1}{c|}{0.0223} & \multicolumn{1}{c|}{1.89} & \multicolumn{1}{c|}{0.2}  & \multicolumn{1}{c|}{0.17} & \multicolumn{1}{c|}{0.08} & \multicolumn{1}{c|}{1.95} & \multicolumn{1}{c|}{0.2}  & \multicolumn{1}{c|}{0.17} & \multicolumn{1}{c|}{0.1}  & \multicolumn{1}{c|}{1.98} & \multicolumn{1}{c|}{0.2}  & \multicolumn{1}{c|}{0.17} & \multicolumn{1}{c|}{0.08} & \multicolumn{1}{c|}{2.03} & \multicolumn{1}{c|}{0.2}  & \multicolumn{1}{c|}{0.17} & \multicolumn{1}{c|}{0.08} \\ \hline
                                   &                             &                             &                           &                           &                           &                           &                           &                           &                           &                           &                           &                           &                           &                           &                           &                           &                           &                           \\ \hline
                                   \multicolumn{1}{|c|}{GC name }    & \multicolumn{1}{c|}{$\sigma_{color} (\Mag_{0})$}  & \multicolumn{1}{c|}{$\sigma_{color} (\Mag_{1})$}&  \multicolumn{1}{c|}{$\alpha$ best fit} & \multicolumn{1}{c|}{$\sigma_{\alpha} (\Mag_{0})$} & \multicolumn{1}{c|}{$\sigma_{\alpha} (\Mag_{1})$} & \multicolumn{1}{c|}{$\sigma_{\alpha} (\Mag_{2})$}  & \multicolumn{1}{c|}{$\alpha$ best fit} & \multicolumn{1}{c|}{$\sigma_{\alpha} (\Mag_{0})$} & \multicolumn{1}{c|}{$\sigma_{\alpha} (\Mag_{1})$} & \multicolumn{1}{c|}{$\sigma_{\alpha} (\Mag_{2})$}  & \multicolumn{1}{c|}{$\alpha$ best fit} & \multicolumn{1}{c|}{$\sigma_{\alpha} (\Mag_{0})$} & \multicolumn{1}{c|}{$\sigma_{\alpha} (\Mag_{1})$} & \multicolumn{1}{c|}{$\sigma_{\alpha} (\Mag_{2})$}   & \multicolumn{1}{c|}{$\alpha$ best fit} & \multicolumn{1}{c|}{$\sigma_{\alpha} (\Mag_{0})$} & \multicolumn{1}{c|}{$\sigma_{\alpha} (\Mag_{1})$} & \multicolumn{1}{c|}{$\sigma_{\alpha} (\Mag_{2})$}     \\ \hline
                                   \multicolumn{1}{|c|}{All 12 GCs}   & \multicolumn{1}{c|}{0.0226} & \multicolumn{1}{c|}{0.0223} & \multicolumn{1}{c|}{1.9} & \multicolumn{1}{c|}{0.21}  & \multicolumn{1}{c|}{0.17} & \multicolumn{1}{c|}{0.1} & \multicolumn{1}{c|}{1.9} & \multicolumn{1}{c|}{0.22}  & \multicolumn{1}{c|}{0.19} & \multicolumn{1}{c|}{0.13}  & \multicolumn{1}{c|}{2} & \multicolumn{1}{c|}{0.2}  & \multicolumn{1}{c|}{0.16} & \multicolumn{1}{c|}{0.1} & \multicolumn{1}{c|}{2} & \multicolumn{1}{c|}{0.2}  & \multicolumn{1}{c|}{0.17} & \multicolumn{1}{c|}{0.12} \\ \hline                               
                                                                      &                             &                             &                           &                           &                           &                           &                           &                           &                           &                           &                           &                           &                           &                           &                           &                           &                           &                           \\ \hline
\multicolumn{19}{|c|}{Sample 2, 11 clusters  with -2.0 \textless {[}Fe/H{]} \textless -1.75}                                                                                                                                                                                                                                                                                                                                                                                                                                                                   \\ \hline
                                   &                             &                             &                           &                           &                           &                           &                           &                           &                           &                           &                           &                           &                           &                           &                           &                           &                           &                           \\ \hline
\multicolumn{1}{|c|}{}             & \multicolumn{1}{c|}{}       & \multicolumn{1}{c|}{}       & \multicolumn{4}{c|}{$Z = 0.00015$}                                                                              & \multicolumn{4}{c|}{$Z = 0.00020$}                                                                              & \multicolumn{4}{c|}{$Z = 0.00025$}                                                                              & \multicolumn{4}{c|}{$Z = 0.00030$}                                                                              \\ \hline
\multicolumn{1}{|c|}{GC name }    & \multicolumn{1}{c|}{$\sigma_{color} (\Mag_{0})$}  & \multicolumn{1}{c|}{$\sigma_{color} (\Mag_{1})$}&  \multicolumn{1}{c|}{$\alpha$ best match} & \multicolumn{1}{c|}{$\sigma_{\alpha} (\Mag_{0})$} & \multicolumn{1}{c|}{$\sigma_{\alpha} (\Mag_{1})$} & \multicolumn{1}{c|}{$\sigma_{\alpha} (\Mag_{2})$}  & \multicolumn{1}{c|}{$\alpha$ best match} & \multicolumn{1}{c|}{$\sigma_{\alpha} (\Mag_{0})$} & \multicolumn{1}{c|}{$\sigma_{\alpha} (\Mag_{1})$} & \multicolumn{1}{c|}{$\sigma_{\alpha} (\Mag_{2})$}  & \multicolumn{1}{c|}{$\alpha$ best match} & \multicolumn{1}{c|}{$\sigma_{\alpha} (\Mag_{0})$} & \multicolumn{1}{c|}{$\sigma_{\alpha} (\Mag_{1})$} & \multicolumn{1}{c|}{$\sigma_{\alpha} (\Mag_{2})$}   & \multicolumn{1}{c|}{$\alpha$ best match} & \multicolumn{1}{c|}{$\sigma_{\alpha} (\Mag_{0})$} & \multicolumn{1}{c|}{$\sigma_{\alpha} (\Mag_{1})$} & \multicolumn{1}{c|}{$\sigma_{\alpha} (\Mag_{2})$}     \\ \hline
\multicolumn{1}{|c|}{Arp 2}        & \multicolumn{1}{c|}{0.0164} & \multicolumn{1}{c|}{0.0156} & \multicolumn{1}{c|}{1.8}  & \multicolumn{1}{c|}{0.16} & \multicolumn{1}{c|}{0.13} & \multicolumn{1}{c|}{N/A}  & \multicolumn{1}{c|}{1.9}  & \multicolumn{1}{c|}{0.15} & \multicolumn{1}{c|}{0.11} & \multicolumn{1}{c|}{N/A}  & \multicolumn{1}{c|}{1.9}  & \multicolumn{1}{c|}{0.15} & \multicolumn{1}{c|}{0.11} & \multicolumn{1}{c|}{N/A}  & \multicolumn{1}{c|}{2}    & \multicolumn{1}{c|}{0.16} & \multicolumn{1}{c|}{0.12} & \multicolumn{1}{c|}{N/A}  \\ \hline
\multicolumn{1}{|c|}{NGC 4147}     & \multicolumn{1}{c|}{0.0155} & \multicolumn{1}{c|}{0.0151} & \multicolumn{1}{c|}{1.8}  & \multicolumn{1}{c|}{0.14} & \multicolumn{1}{c|}{0.12} & \multicolumn{1}{c|}{N/A}  & \multicolumn{1}{c|}{1.8}  & \multicolumn{1}{c|}{0.14} & \multicolumn{1}{c|}{0.12} & \multicolumn{1}{c|}{N/A}  & \multicolumn{1}{c|}{1.9}  & \multicolumn{1}{c|}{0.16} & \multicolumn{1}{c|}{0.11} & \multicolumn{1}{c|}{N/A}  & \multicolumn{1}{c|}{1.9}  & \multicolumn{1}{c|}{0.15} & \multicolumn{1}{c|}{0.11} & \multicolumn{1}{c|}{N/A}  \\ \hline
\multicolumn{1}{|c|}{NGC 5024}     & \multicolumn{1}{c|}{0.0173} & \multicolumn{1}{c|}{0.0193} & \multicolumn{1}{c|}{1.8}  & \multicolumn{1}{c|}{0.17} & \multicolumn{1}{c|}{0.16} & \multicolumn{1}{c|}{0.07} & \multicolumn{1}{c|}{1.9}  & \multicolumn{1}{c|}{0.16} & \multicolumn{1}{c|}{0.14} & \multicolumn{1}{c|}{0.05} & \multicolumn{1}{c|}{1.9}  & \multicolumn{1}{c|}{0.16} & \multicolumn{1}{c|}{0.14} & \multicolumn{1}{c|}{0.06} & \multicolumn{1}{c|}{2}    & \multicolumn{1}{c|}{0.16} & \multicolumn{1}{c|}{0.14} & \multicolumn{1}{c|}{0.06} \\ \hline
\multicolumn{1}{|c|}{NGC 5466}     & \multicolumn{1}{c|}{0.0154} & \multicolumn{1}{c|}{0.0159} & \multicolumn{1}{c|}{1.8}  & \multicolumn{1}{c|}{0.14} & \multicolumn{1}{c|}{0.12} & \multicolumn{1}{c|}{N/A}  & \multicolumn{1}{c|}{1.9}  & \multicolumn{1}{c|}{0.14} & \multicolumn{1}{c|}{0.11} & \multicolumn{1}{c|}{N/A}  & \multicolumn{1}{c|}{1.9}  & \multicolumn{1}{c|}{0.13} & \multicolumn{1}{c|}{0.11} & \multicolumn{1}{c|}{N/A}  & \multicolumn{1}{c|}{1.9}  & \multicolumn{1}{c|}{0.13} & \multicolumn{1}{c|}{0.11} & \multicolumn{1}{c|}{N/A}  \\ \hline
\multicolumn{1}{|c|}{NGC 6093}     & \multicolumn{1}{c|}{0.0212} & \multicolumn{1}{c|}{0.022}  & \multicolumn{1}{c|}{1.8}  & \multicolumn{1}{c|}{0.21} & \multicolumn{1}{c|}{0.16} & \multicolumn{1}{c|}{0.08} & \multicolumn{1}{c|}{1.8}  & \multicolumn{1}{c|}{0.21} & \multicolumn{1}{c|}{0.16} & \multicolumn{1}{c|}{0.1}  & \multicolumn{1}{c|}{1.9}  & \multicolumn{1}{c|}{0.2}  & \multicolumn{1}{c|}{0.14} & \multicolumn{1}{c|}{0.08} & \multicolumn{1}{c|}{1.9}  & \multicolumn{1}{c|}{0.2}  & \multicolumn{1}{c|}{0.15} & \multicolumn{1}{c|}{0.09} \\ \hline
\multicolumn{1}{|c|}{NGC 6101}     & \multicolumn{1}{c|}{0.0181} & \multicolumn{1}{c|}{0.0215} & \multicolumn{1}{c|}{1.8}  & \multicolumn{1}{c|}{0.15} & \multicolumn{1}{c|}{0.15} & \multicolumn{1}{c|}{0.07} & \multicolumn{1}{c|}{1.8}  & \multicolumn{1}{c|}{0.16} & \multicolumn{1}{c|}{0.16} & \multicolumn{1}{c|}{0.1}  & \multicolumn{1}{c|}{1.9}  & \multicolumn{1}{c|}{0.16} & \multicolumn{1}{c|}{0.14} & \multicolumn{1}{c|}{0.06} & \multicolumn{1}{c|}{1.9}  & \multicolumn{1}{c|}{0.16} & \multicolumn{1}{c|}{0.14} & \multicolumn{1}{c|}{0.08} \\ \hline
\multicolumn{1}{|c|}{NGC 6144}     & \multicolumn{1}{c|}{0.027}  & \multicolumn{1}{c|}{0.0308} & \multicolumn{1}{c|}{1.8}  & \multicolumn{1}{c|}{0.24} & \multicolumn{1}{c|}{0.21} & \multicolumn{1}{c|}{N/A}  & \multicolumn{1}{c|}{1.8}  & \multicolumn{1}{c|}{0.25} & \multicolumn{1}{c|}{0.22} & \multicolumn{1}{c|}{N/A}  & \multicolumn{1}{c|}{1.8}  & \multicolumn{1}{c|}{0.26} & \multicolumn{1}{c|}{0.24} & \multicolumn{1}{c|}{N/A}  & \multicolumn{1}{c|}{1.9}  & \multicolumn{1}{c|}{0.22} & \multicolumn{1}{c|}{0.19} & \multicolumn{1}{c|}{N/A}  \\ \hline
\multicolumn{1}{|c|}{NGC 6254}     & \multicolumn{1}{c|}{0.0226} & \multicolumn{1}{c|}{0.0219} & \multicolumn{1}{c|}{1.8}  & \multicolumn{1}{c|}{0.24} & \multicolumn{1}{c|}{0.15} & \multicolumn{1}{c|}{0.08} & \multicolumn{1}{c|}{1.9}  & \multicolumn{1}{c|}{0.21} & \multicolumn{1}{c|}{0.14} & \multicolumn{1}{c|}{0.09} & \multicolumn{1}{c|}{1.9}  & \multicolumn{1}{c|}{0.22} & \multicolumn{1}{c|}{0.14} & \multicolumn{1}{c|}{0.08} & \multicolumn{1}{c|}{1.9}  & \multicolumn{1}{c|}{0.22} & \multicolumn{1}{c|}{0.14} & \multicolumn{1}{c|}{0.08} \\ \hline
\multicolumn{1}{|c|}{NGC 6535}     & \multicolumn{1}{c|}{0.0201} & \multicolumn{1}{c|}{0.0155} & \multicolumn{1}{c|}{1.8}  & \multicolumn{1}{c|}{0.21} & \multicolumn{1}{c|}{0.09} & \multicolumn{1}{c|}{N/A}  & \multicolumn{1}{c|}{1.9}  & \multicolumn{1}{c|}{0.17} & \multicolumn{1}{c|}{0.07} & \multicolumn{1}{c|}{N/A}  & \multicolumn{1}{c|}{1.9}  & \multicolumn{1}{c|}{0.17} & \multicolumn{1}{c|}{0.07} & \multicolumn{1}{c|}{N/A}  & \multicolumn{1}{c|}{1.9}  & \multicolumn{1}{c|}{0.18} & \multicolumn{1}{c|}{0.08} & \multicolumn{1}{c|}{N/A}  \\ \hline
\multicolumn{1}{|c|}{NGC 6541}     & \multicolumn{1}{c|}{0.0216} & \multicolumn{1}{c|}{0.0208} & \multicolumn{1}{c|}{1.9}  & \multicolumn{1}{c|}{0.19} & \multicolumn{1}{c|}{0.14} & \multicolumn{1}{c|}{0.07} & \multicolumn{1}{c|}{1.9}  & \multicolumn{1}{c|}{0.2}  & \multicolumn{1}{c|}{0.14} & \multicolumn{1}{c|}{0.07} & \multicolumn{1}{c|}{2}    & \multicolumn{1}{c|}{0.18} & \multicolumn{1}{c|}{0.14} & \multicolumn{1}{c|}{0.08} & \multicolumn{1}{c|}{2}    & \multicolumn{1}{c|}{0.18} & \multicolumn{1}{c|}{0.13} & \multicolumn{1}{c|}{0.07} \\ \hline
\multicolumn{1}{|c|}{NGC 6809}     & \multicolumn{1}{c|}{0.0177} & \multicolumn{1}{c|}{0.02}   & \multicolumn{1}{c|}{1.8}  & \multicolumn{1}{c|}{0.15} & \multicolumn{1}{c|}{0.15} & \multicolumn{1}{c|}{N/A}  & \multicolumn{1}{c|}{1.8}  & \multicolumn{1}{c|}{0.15} & \multicolumn{1}{c|}{0.15} & \multicolumn{1}{c|}{N/A}  & \multicolumn{1}{c|}{1.8}  & \multicolumn{1}{c|}{0.15} & \multicolumn{1}{c|}{0.15} & \multicolumn{1}{c|}{N/A}  & \multicolumn{1}{c|}{1.9}  & \multicolumn{1}{c|}{0.17} & \multicolumn{1}{c|}{0.14} & \multicolumn{1}{c|}{N/A}  \\ \hline
\multicolumn{1}{|c|}{All 11 GCs}   & \multicolumn{1}{c|}{0.0205} & \multicolumn{1}{c|}{0.0217} & \multicolumn{1}{c|}{1.81} & \multicolumn{1}{c|}{0.18} & \multicolumn{1}{c|}{0.14} & \multicolumn{1}{c|}{0.07} & \multicolumn{1}{c|}{1.85} & \multicolumn{1}{c|}{0.18} & \multicolumn{1}{c|}{0.14} & \multicolumn{1}{c|}{0.08} & \multicolumn{1}{c|}{1.89} & \multicolumn{1}{c|}{0.18} & \multicolumn{1}{c|}{0.14} & \multicolumn{1}{c|}{0.07} & \multicolumn{1}{c|}{1.93} & \multicolumn{1}{c|}{0.18} & \multicolumn{1}{c|}{0.13} & \multicolumn{1}{c|}{0.08} \\ \hline
                                   &                             &                             &                           &                           &                           &                           &                           &                           &                           &                           &                           &                           &                           &                           &                           &                           &                           &                           \\ \hline
                                                                      \multicolumn{1}{|c|}{GC name }    & \multicolumn{1}{c|}{$\sigma_{color} (\Mag_{0})$}  & \multicolumn{1}{c|}{$\sigma_{color} (\Mag_{1})$}&  \multicolumn{1}{c|}{$\alpha$ best fit} & \multicolumn{1}{c|}{$\sigma_{\alpha} (\Mag_{0})$} & \multicolumn{1}{c|}{$\sigma_{\alpha} (\Mag_{1})$} & \multicolumn{1}{c|}{$\sigma_{\alpha} (\Mag_{2})$}  & \multicolumn{1}{c|}{$\alpha$ best fit} & \multicolumn{1}{c|}{$\sigma_{\alpha} (\Mag_{0})$} & \multicolumn{1}{c|}{$\sigma_{\alpha} (\Mag_{1})$} & \multicolumn{1}{c|}{$\sigma_{\alpha} (\Mag_{2})$}  & \multicolumn{1}{c|}{$\alpha$ best fit} & \multicolumn{1}{c|}{$\sigma_{\alpha} (\Mag_{0})$} & \multicolumn{1}{c|}{$\sigma_{\alpha} (\Mag_{1})$} & \multicolumn{1}{c|}{$\sigma_{\alpha} (\Mag_{2})$}   & \multicolumn{1}{c|}{$\alpha$ best fit} & \multicolumn{1}{c|}{$\sigma_{\alpha} (\Mag_{0})$} & \multicolumn{1}{c|}{$\sigma_{\alpha} (\Mag_{1})$} & \multicolumn{1}{c|}{$\sigma_{\alpha} (\Mag_{2})$}     \\ \hline
\multicolumn{1}{|c|}{All 11 GCs}   & \multicolumn{1}{c|}{0.0205} & \multicolumn{1}{c|}{0.0217} & \multicolumn{1}{c|}{1.8} & \multicolumn{1}{c|}{0.2} & \multicolumn{1}{c|}{0.16} & \multicolumn{1}{c|}{0.09} & \multicolumn{1}{c|}{1.9} & \multicolumn{1}{c|}{0.19} & \multicolumn{1}{c|}{0.14} & \multicolumn{1}{c|}{0.07} & \multicolumn{1}{c|}{1.9} & \multicolumn{1}{c|}{0.19} & \multicolumn{1}{c|}{0.14} & \multicolumn{1}{c|}{0.07} & \multicolumn{1}{c|}{1.9} & \multicolumn{1}{c|}{0.19} & \multicolumn{1}{c|}{0.15} & \multicolumn{1}{c|}{0.09} \\ \hline                                                                   
                                    
                                                                      &                             &                             &                           &                           &                           &                           &                           &                           &                           &                           &                           &                           &                           &                           &                           &                           &                           &                           \\ \hline
\multicolumn{19}{|c|}{Sample 3, 15 clusters  with -1.75 \textless {[}Fe/H{]} \textless -1.5}                                                                                                                                                                                                                                                                                                                                                                                                                                                                   \\ \hline
                                   &                             &                             &                           &                           &                           &                           &                           &                           &                           &                           &                           &                           &                           &                           &                           &                           &                           &                           \\ \hline
\multicolumn{1}{|c|}{}             & \multicolumn{1}{c|}{}       & \multicolumn{1}{c|}{}       & \multicolumn{4}{c|}{$Z = 0.00025$}                                                                              & \multicolumn{4}{c|}{$Z = 0.00030$}                                                                              & \multicolumn{4}{c|}{$Z = 0.00035$}                                                                              & \multicolumn{4}{c|}{$Z = 0.00040$}                                                                              \\ \hline
\multicolumn{1}{|c|}{GC name }    & \multicolumn{1}{c|}{$\sigma_{color} (\Mag_{0})$}  & \multicolumn{1}{c|}{$\sigma_{color} (\Mag_{1})$}&  \multicolumn{1}{c|}{$\alpha$ best match} & \multicolumn{1}{c|}{$\sigma_{\alpha} (\Mag_{0})$} & \multicolumn{1}{c|}{$\sigma_{\alpha} (\Mag_{1})$} & \multicolumn{1}{c|}{$\sigma_{\alpha} (\Mag_{2})$}  & \multicolumn{1}{c|}{$\alpha$ best match} & \multicolumn{1}{c|}{$\sigma_{\alpha} (\Mag_{0})$} & \multicolumn{1}{c|}{$\sigma_{\alpha} (\Mag_{1})$} & \multicolumn{1}{c|}{$\sigma_{\alpha} (\Mag_{2})$}  & \multicolumn{1}{c|}{$\alpha$ best match} & \multicolumn{1}{c|}{$\sigma_{\alpha} (\Mag_{0})$} & \multicolumn{1}{c|}{$\sigma_{\alpha} (\Mag_{1})$} & \multicolumn{1}{c|}{$\sigma_{\alpha} (\Mag_{2})$}   & \multicolumn{1}{c|}{$\alpha$ best match} & \multicolumn{1}{c|}{$\sigma_{\alpha} (\Mag_{0})$} & \multicolumn{1}{c|}{$\sigma_{\alpha} (\Mag_{1})$} & \multicolumn{1}{c|}{$\sigma_{\alpha} (\Mag_{2})$}     \\ \hline
\multicolumn{1}{|c|}{IC4499}       & \multicolumn{1}{c|}{0.0186} & \multicolumn{1}{c|}{0.0215} & \multicolumn{1}{c|}{1.9}  & \multicolumn{1}{c|}{0.16} & \multicolumn{1}{c|}{0.14} & \multicolumn{1}{c|}{N/A}  & \multicolumn{1}{c|}{1.9}  & \multicolumn{1}{c|}{0.16} & \multicolumn{1}{c|}{0.14} & \multicolumn{1}{c|}{N/A}  & \multicolumn{1}{c|}{1.9}  & \multicolumn{1}{c|}{0.16} & \multicolumn{1}{c|}{0.15} & \multicolumn{1}{c|}{N/A}  & \multicolumn{1}{c|}{2}    & \multicolumn{1}{c|}{0.16} & \multicolumn{1}{c|}{0.13} & \multicolumn{1}{c|}{N/A}  \\ \hline
\multicolumn{1}{|c|}{NGC 3201}     & \multicolumn{1}{c|}{0.0241} & \multicolumn{1}{c|}{0.0248} & \multicolumn{1}{c|}{1.8}  & \multicolumn{1}{c|}{0.18} & \multicolumn{1}{c|}{0.14} & \multicolumn{1}{c|}{0.08} & \multicolumn{1}{c|}{1.8}  & \multicolumn{1}{c|}{0.17} & \multicolumn{1}{c|}{0.13} & \multicolumn{1}{c|}{0.06} & \multicolumn{1}{c|}{1.8}  & \multicolumn{1}{c|}{0.18} & \multicolumn{1}{c|}{0.13} & \multicolumn{1}{c|}{0.07} & \multicolumn{1}{c|}{1.9}  & \multicolumn{1}{c|}{0.2}  & \multicolumn{1}{c|}{0.14} & \multicolumn{1}{c|}{0.08} \\ \hline
\multicolumn{1}{|c|}{NGC 5139}     & \multicolumn{1}{c|}{0.0192} & \multicolumn{1}{c|}{0.0211} & \multicolumn{1}{c|}{1.9}  & \multicolumn{1}{c|}{0.17} & \multicolumn{1}{c|}{0.14} & \multicolumn{1}{c|}{0.06} & \multicolumn{1}{c|}{1.9}  & \multicolumn{1}{c|}{0.17} & \multicolumn{1}{c|}{0.14} & \multicolumn{1}{c|}{0.07} & \multicolumn{1}{c|}{1.9}  & \multicolumn{1}{c|}{0.17} & \multicolumn{1}{c|}{0.14} & \multicolumn{1}{c|}{0.09} & \multicolumn{1}{c|}{2}    & \multicolumn{1}{c|}{0.18} & \multicolumn{1}{c|}{0.14} & \multicolumn{1}{c|}{0.06} \\ \hline
\multicolumn{1}{|c|}{NGC 5272}     & \multicolumn{1}{c|}{0.0157} & \multicolumn{1}{c|}{0.0202} & \multicolumn{1}{c|}{1.7}  & \multicolumn{1}{c|}{0.15} & \multicolumn{1}{c|}{0.14} & \multicolumn{1}{c|}{0.08} & \multicolumn{1}{c|}{1.8}  & \multicolumn{1}{c|}{0.13} & \multicolumn{1}{c|}{0.13} & \multicolumn{1}{c|}{0.05} & \multicolumn{1}{c|}{1.8}  & \multicolumn{1}{c|}{0.13} & \multicolumn{1}{c|}{0.13} & \multicolumn{1}{c|}{0.06} & \multicolumn{1}{c|}{1.8}  & \multicolumn{1}{c|}{0.14} & \multicolumn{1}{c|}{0.13} & \multicolumn{1}{c|}{0.07} \\ \hline
\multicolumn{1}{|c|}{NGC 5286}     & \multicolumn{1}{c|}{0.0221} & \multicolumn{1}{c|}{0.0228} & \multicolumn{1}{c|}{1.9}  & \multicolumn{1}{c|}{0.2}  & \multicolumn{1}{c|}{0.16} & \multicolumn{1}{c|}{0.07} & \multicolumn{1}{c|}{1.9}  & \multicolumn{1}{c|}{0.21} & \multicolumn{1}{c|}{0.17} & \multicolumn{1}{c|}{0.09} & \multicolumn{1}{c|}{2}    & \multicolumn{1}{c|}{0.18} & \multicolumn{1}{c|}{0.15} & \multicolumn{1}{c|}{0.06} & \multicolumn{1}{c|}{2}    & \multicolumn{1}{c|}{0.18} & \multicolumn{1}{c|}{0.15} & \multicolumn{1}{c|}{0.07} \\ \hline
\multicolumn{1}{|c|}{NGC 5986}     & \multicolumn{1}{c|}{0.0268} & \multicolumn{1}{c|}{0.0285} & \multicolumn{1}{c|}{1.8}  & \multicolumn{1}{c|}{0.23} & \multicolumn{1}{c|}{0.17} & \multicolumn{1}{c|}{0.07} & \multicolumn{1}{c|}{1.8}  & \multicolumn{1}{c|}{0.23} & \multicolumn{1}{c|}{0.19} & \multicolumn{1}{c|}{0.08} & \multicolumn{1}{c|}{1.9}  & \multicolumn{1}{c|}{0.22} & \multicolumn{1}{c|}{0.16} & \multicolumn{1}{c|}{0.08} & \multicolumn{1}{c|}{1.9}  & \multicolumn{1}{c|}{0.22} & \multicolumn{1}{c|}{0.16} & \multicolumn{1}{c|}{0.07} \\ \hline
\multicolumn{1}{|c|}{NGC 6218}     & \multicolumn{1}{c|}{0.0161} & \multicolumn{1}{c|}{0.0141} & \multicolumn{1}{c|}{1.7}  & \multicolumn{1}{c|}{0.15} & \multicolumn{1}{c|}{0.07} & \multicolumn{1}{c|}{N/A}  & \multicolumn{1}{c|}{1.7}  & \multicolumn{1}{c|}{0.15} & \multicolumn{1}{c|}{0.07} & \multicolumn{1}{c|}{N/A}  & \multicolumn{1}{c|}{1.8}  & \multicolumn{1}{c|}{0.12} & \multicolumn{1}{c|}{0.08} & \multicolumn{1}{c|}{N/A}  & \multicolumn{1}{c|}{1.8}  & \multicolumn{1}{c|}{0.12} & \multicolumn{1}{c|}{0.07} & \multicolumn{1}{c|}{N/A}  \\ \hline
\multicolumn{1}{|c|}{NGC 6584}     & \multicolumn{1}{c|}{0.0173} & \multicolumn{1}{c|}{0.0194} & \multicolumn{1}{c|}{1.8}  & \multicolumn{1}{c|}{0.16} & \multicolumn{1}{c|}{0.14} & \multicolumn{1}{c|}{0.05} & \multicolumn{1}{c|}{1.8}  & \multicolumn{1}{c|}{0.17} & \multicolumn{1}{c|}{0.15} & \multicolumn{1}{c|}{0.06} & \multicolumn{1}{c|}{1.9}  & \multicolumn{1}{c|}{0.14} & \multicolumn{1}{c|}{0.13} & \multicolumn{1}{c|}{0.06} & \multicolumn{1}{c|}{1.9}  & \multicolumn{1}{c|}{0.14} & \multicolumn{1}{c|}{0.13} & \multicolumn{1}{c|}{0.05} \\ \hline
\multicolumn{1}{|c|}{NGC 6656}     & \multicolumn{1}{c|}{0.0233} & \multicolumn{1}{c|}{0.0243} & \multicolumn{1}{c|}{1.9}  & \multicolumn{1}{c|}{0.19} & \multicolumn{1}{c|}{0.14} & \multicolumn{1}{c|}{0.07} & \multicolumn{1}{c|}{1.9}  & \multicolumn{1}{c|}{0.2}  & \multicolumn{1}{c|}{0.15} & \multicolumn{1}{c|}{0.1}  & \multicolumn{1}{c|}{2}    & \multicolumn{1}{c|}{0.17} & \multicolumn{1}{c|}{0.13} & \multicolumn{1}{c|}{0.05} & \multicolumn{1}{c|}{2}    & \multicolumn{1}{c|}{0.17} & \multicolumn{1}{c|}{0.13} & \multicolumn{1}{c|}{0.07} \\ \hline
\multicolumn{1}{|c|}{NGC 6681}     & \multicolumn{1}{c|}{0.0206} & \multicolumn{1}{c|}{0.0214} & \multicolumn{1}{c|}{1.8}  & \multicolumn{1}{c|}{0.19} & \multicolumn{1}{c|}{0.15} & \multicolumn{1}{c|}{0.09} & \multicolumn{1}{c|}{1.9}  & \multicolumn{1}{c|}{0.18} & \multicolumn{1}{c|}{0.14} & \multicolumn{1}{c|}{0.1}  & \multicolumn{1}{c|}{1.9}  & \multicolumn{1}{c|}{0.18} & \multicolumn{1}{c|}{0.13} & \multicolumn{1}{c|}{0.09} & \multicolumn{1}{c|}{1.9}  & \multicolumn{1}{c|}{0.18} & \multicolumn{1}{c|}{0.14} & \multicolumn{1}{c|}{0.09} \\ \hline
\multicolumn{1}{|c|}{NGC 6752}     & \multicolumn{1}{c|}{0.0174} & \multicolumn{1}{c|}{0.0204} & \multicolumn{1}{c|}{1.7}  & \multicolumn{1}{c|}{0.17} & \multicolumn{1}{c|}{0.12} & \multicolumn{1}{c|}{0.11} & \multicolumn{1}{c|}{1.8}  & \multicolumn{1}{c|}{0.14} & \multicolumn{1}{c|}{0.11} & \multicolumn{1}{c|}{0.08} & \multicolumn{1}{c|}{1.8}  & \multicolumn{1}{c|}{0.14} & \multicolumn{1}{c|}{0.11} & \multicolumn{1}{c|}{0.09} & \multicolumn{1}{c|}{1.8}  & \multicolumn{1}{c|}{0.15} & \multicolumn{1}{c|}{0.11} & \multicolumn{1}{c|}{0.11} \\ \hline
\multicolumn{1}{|c|}{NGC 6934}     & \multicolumn{1}{c|}{0.017}  & \multicolumn{1}{c|}{0.0211} & \multicolumn{1}{c|}{1.8}  & \multicolumn{1}{c|}{0.15} & \multicolumn{1}{c|}{0.14} & \multicolumn{1}{c|}{0.08} & \multicolumn{1}{c|}{1.8}  & \multicolumn{1}{c|}{0.15} & \multicolumn{1}{c|}{0.14} & \multicolumn{1}{c|}{0.06} & \multicolumn{1}{c|}{1.8}  & \multicolumn{1}{c|}{0.16} & \multicolumn{1}{c|}{0.14} & \multicolumn{1}{c|}{0.06} & \multicolumn{1}{c|}{1.9}  & \multicolumn{1}{c|}{0.14} & \multicolumn{1}{c|}{0.13} & \multicolumn{1}{c|}{0.08} \\ \hline
\multicolumn{1}{|c|}{NGC 7006}     & \multicolumn{1}{c|}{0.0197} & \multicolumn{1}{c|}{0.0258} & \multicolumn{1}{c|}{1.8}  & \multicolumn{1}{c|}{0.17} & \multicolumn{1}{c|}{0.18} & \multicolumn{1}{c|}{0.09} & \multicolumn{1}{c|}{1.9}  & \multicolumn{1}{c|}{0.17} & \multicolumn{1}{c|}{0.17} & \multicolumn{1}{c|}{0.1}  & \multicolumn{1}{c|}{1.9}  & \multicolumn{1}{c|}{0.17} & \multicolumn{1}{c|}{0.17} & \multicolumn{1}{c|}{0.09} & \multicolumn{1}{c|}{1.9}  & \multicolumn{1}{c|}{0.17} & \multicolumn{1}{c|}{0.17} & \multicolumn{1}{c|}{0.09} \\ \hline
\multicolumn{1}{|c|}{NGC 7089}     & \multicolumn{1}{c|}{0.0176} & \multicolumn{1}{c|}{0.0184} & \multicolumn{1}{c|}{1.8}  & \multicolumn{1}{c|}{0.18} & \multicolumn{1}{c|}{0.14} & \multicolumn{1}{c|}{0.05} & \multicolumn{1}{c|}{1.9}  & \multicolumn{1}{c|}{0.16} & \multicolumn{1}{c|}{0.13} & \multicolumn{1}{c|}{0.06} & \multicolumn{1}{c|}{1.9}  & \multicolumn{1}{c|}{0.16} & \multicolumn{1}{c|}{0.13} & \multicolumn{1}{c|}{0.05} & \multicolumn{1}{c|}{1.9}  & \multicolumn{1}{c|}{0.16} & \multicolumn{1}{c|}{0.13} & \multicolumn{1}{c|}{0.05} \\ \hline
\multicolumn{1}{|c|}{Ruprecht 106} & \multicolumn{1}{c|}{0.0223} & \multicolumn{1}{c|}{0.0202} & \multicolumn{1}{c|}{1.9}  & \multicolumn{1}{c|}{0.22} & \multicolumn{1}{c|}{0.13} & \multicolumn{1}{c|}{N/A}  & \multicolumn{1}{c|}{1.9}  & \multicolumn{1}{c|}{0.23} & \multicolumn{1}{c|}{0.14} & \multicolumn{1}{c|}{N/A}  & \multicolumn{1}{c|}{2}    & \multicolumn{1}{c|}{0.18} & \multicolumn{1}{c|}{0.11} & \multicolumn{1}{c|}{N/A}  & \multicolumn{1}{c|}{2}    & \multicolumn{1}{c|}{0.19} & \multicolumn{1}{c|}{0.11} & \multicolumn{1}{c|}{N/A}  \\ \hline
\multicolumn{1}{|c|}{All 15 GCs}   & \multicolumn{1}{c|}{0.021} & \multicolumn{1}{c|}{0.0251} & \multicolumn{1}{c|}{1.81} & \multicolumn{1}{c|}{0.18} & \multicolumn{1}{c|}{0.14} & \multicolumn{1}{c|}{0.07} & \multicolumn{1}{c|}{1.85} & \multicolumn{1}{c|}{0.17} & \multicolumn{1}{c|}{0.14} & \multicolumn{1}{c|}{0.08} & \multicolumn{1}{c|}{1.89} & \multicolumn{1}{c|}{0.16} & \multicolumn{1}{c|}{0.13} & \multicolumn{1}{c|}{0.07} & \multicolumn{1}{c|}{1.91} & \multicolumn{1}{c|}{0.17} & \multicolumn{1}{c|}{0.13} & \multicolumn{1}{c|}{0.07} \\ \hline
                                                                      &                             &                             &                           &                           &                           &                           &                           &                           &                           &                           &                           &                           &                           &                           &                           &                           &                           &                           \\ \hline
\multicolumn{1}{|c|}{GC name }    & \multicolumn{1}{c|}{$\sigma_{color} (\Mag_{0})$}  & \multicolumn{1}{c|}{$\sigma_{color} (\Mag_{1})$}&  \multicolumn{1}{c|}{$\alpha$ best fit} & \multicolumn{1}{c|}{$\sigma_{\alpha} (\Mag_{0})$} & \multicolumn{1}{c|}{$\sigma_{\alpha} (\Mag_{1})$} & \multicolumn{1}{c|}{$\sigma_{\alpha} (\Mag_{2})$}  & \multicolumn{1}{c|}{$\alpha$ best fit} & \multicolumn{1}{c|}{$\sigma_{\alpha} (\Mag_{0})$} & \multicolumn{1}{c|}{$\sigma_{\alpha} (\Mag_{1})$} & \multicolumn{1}{c|}{$\sigma_{\alpha} (\Mag_{2})$}  & \multicolumn{1}{c|}{$\alpha$ best fit} & \multicolumn{1}{c|}{$\sigma_{\alpha} (\Mag_{0})$} & \multicolumn{1}{c|}{$\sigma_{\alpha} (\Mag_{1})$} & \multicolumn{1}{c|}{$\sigma_{\alpha} (\Mag_{2})$}   & \multicolumn{1}{c|}{$\alpha$ best fit} & \multicolumn{1}{c|}{$\sigma_{\alpha} (\Mag_{0})$} & \multicolumn{1}{c|}{$\sigma_{\alpha} (\Mag_{1})$} & \multicolumn{1}{c|}{$\sigma_{\alpha} (\Mag_{2})$}     \\ \hline
\multicolumn{1}{|c|}{All 15 GCs}   & \multicolumn{1}{c|}{0.021} & \multicolumn{1}{c|}{0.0251} & \multicolumn{1}{c|}{1.8} & \multicolumn{1}{c|}{0.19} & \multicolumn{1}{c|}{0.16} & \multicolumn{1}{c|}{0.1} & \multicolumn{1}{c|}{1.8} & \multicolumn{1}{c|}{0.19} & \multicolumn{1}{c|}{0.17} & \multicolumn{1}{c|}{0.12} & \multicolumn{1}{c|}{1.9} & \multicolumn{1}{c|}{0.18} & \multicolumn{1}{c|}{0.15} & \multicolumn{1}{c|}{0.1} & \multicolumn{1}{c|}{1.9} & \multicolumn{1}{c|}{0.18} & \multicolumn{1}{c|}{0.15} & \multicolumn{1}{c|}{0.1} \\ \hline
         
                                                                      &                             &                             &                           &                           &                           &                           &                           &                           &                           &                           &                           &                           &                           &                           &                           &                           &                           &                           \\ \hline
\end{tabular}%
}
\caption{Dispersion in color and mixing length for the 4 fiducial adopted metallicity values. The results are given for each cluster in each of the sub-samples and all  combined. For the combined sample we report both the means of the best match $\alpha_{\rm MLT}$ values and the best-fit $\alpha_{\rm MLT}$.   For magnitude bins with less than 10 stars, the dispersion  $\alpha$ is not computed (N/A).  The color dispersion is not affected by the choice of initial metallicity. }
\label{table:alldispersion}
\end{table}

\begin{figure}
\centering
\includegraphics[width=0.8\textwidth]{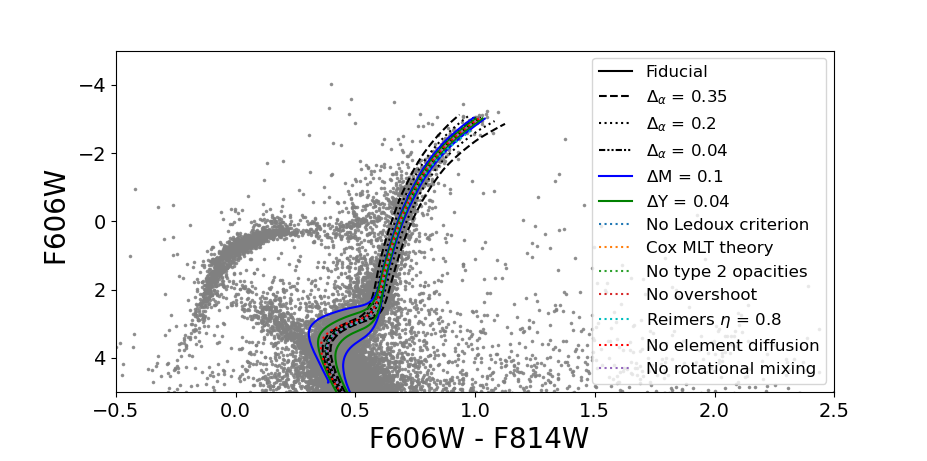} 
\includegraphics[width=0.8\textwidth]{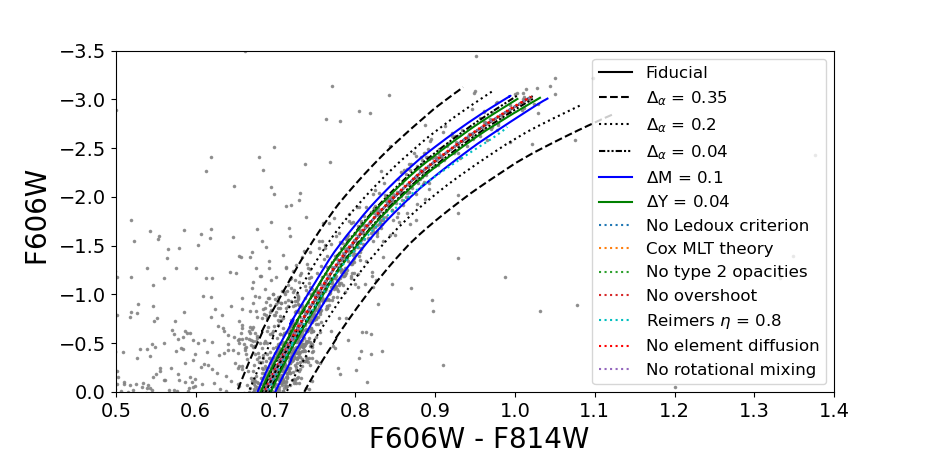} 
\includegraphics[width=0.8\textwidth]{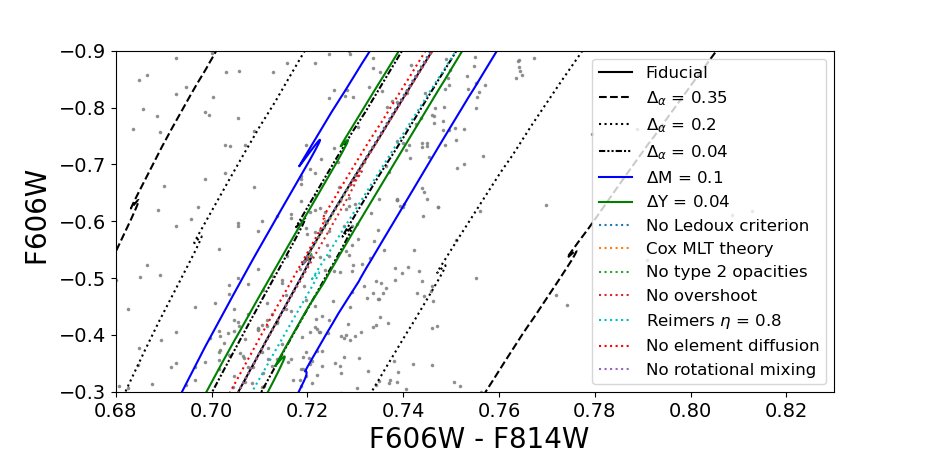} 
\caption{The effect on the RGB of  (extreme) changes of other microphysics parameters besides the mixing length. The three panels are subsequent zoom ins in the relevant part of the CMD.   Note that the effect is much smaller  than that of $\alpha_{\rm MLT}$, and well below  the intrinsic broadening of the RGB (gray points) and thus below star-to-star color variations.}
\label{fig:otherparam}
\end{figure}

\section{Assessing robustness: Tests of microphysics}
\label{sec:robustnesstests}
Throughout this work, we have assumed a fixed $\alpha_{\rm MLT}$ through the stellar tracks: this value does not change in itself the duration of the lifetime of the star. The lifetime of low-mass stars to helium flash is mostly dependent on mass and metallicity. In fact, the lifetime in Gyr to the He flash can be fitted with 5\% precision by the formula (adopting a helium mass fraction $Y=0.24$)
\begin{equation}
\log t = -0.207-3.691 \log (M/M_{\odot}) +11.327 \log (1.76-Z) +0.870 \log (0.0086+Z) 
\end{equation}

Beside the mixing length and the metallicity, which, as we have seen,  determine the color of the RGB to leading  and next-to-leading order, other parameters can also affect the color of the RGB, although with a weaker dependence:  initial mass, initial helium mass fraction, overshooting of the convection depth, mass loss, rotational mixing and element diffusion. These are kept fixed at their fiducial values in our main analysis, here we quantify their effect. We also explore changing the mixing length theory to the Cox formulation \cite{Cox} .    
 We compute stellar tracks by varying a single parameter  at the time  while keeping the other parameters  fixed to the fiducial configuration (see section \ref{tracks}) with  a mixing length value $\alpha = 1.9$ being this the closest  in our grid to the solar value  1.938.

This is illustrated in Figure \ref{fig:otherparam} where the top panel shows the full RGB and the bottom panel is a zoom-in on the Helium flash region, to make visible the (small) effect of some of the parameters we consider.
Variations of initial mass,  initial helium mass fraction  and  diffusion,  which alters the lifetime of the star, affect  the main sequence more severely than the RGB. The rest of the parameters  influence the more advanced life stages of the star. Type 2 opacities are important to compute the helium burning rate  and a variation of the mass loss parameter cause the tracks to deviate towards the end of the RGB. 
 It is well known that a shift  in mass of  $\Delta M=0.1$ or a shift  in Helium  of  $\Delta Y=0.04$,  are heavily disfavoured by the data as  these parameters also change the main sequence and the MSTO and  with such shifts   models do not fit this part of the CMD. We also test different models of convection (Henyey and Cox \cite{Cox,Henyey} with Henyey parameters y and $\nu$ respectively defined to 1/3 and 8.), the implementation or not of overshoot (we use an exponential scheme for every zone of the core $f_{\rm core} = 0.016$ and the envelope $f_{\rm core} = 0.0174$, similar to Ref.~\cite{MIST1}), and with or without Ledoux criterion (the Ledoux criterion is associated to a semiconvection factor $\alpha_{sc} = 0.1$ and when it is not activated the default criterion is reverted back to  Schwarzschild. For more information see Ref. [6] and references therein.). The value for the mass loss Reimers parameter $\eta=0.8$ is an extreme value motivated by the morphology the horizontal branch of  the CMD of globular clusters (e.g.,~\cite{JimenezGC96,etaref}).  For completeness, we also show rotational mixing, even though it mainly affects massive stars.  This is the motivation for the  adopted choices in the Figure. 
Most of the parameters tested have little to no impact for low mass stars or will affect the horizontal or asymptotic branch which are not of interest for this paper. 

Changes in the initial mass $\Delta M$ and in the initial helium mass fraction $\Delta Y$ have a much bigger effect than all other changes, which are much more subtle, and small compared to the  intrinsic broadening of the RGB.
 Unlike $\alpha_{\rm LMT}$, the other parameters of the mixing length theory do modify the stellar tracks  but maintaining the color of the RGB unaltered. 
We also illustrate the effect of  $\Delta_{\alpha}=0.35$ , the value very conservatively adopted by \cite{ValcinGC}, and   $\Delta_{\alpha}=0.2$ and $0.04$ as estimated by the two approaches presented in the main text. These considerations indicate that the results reported are robust to  any possible changes in key stellar  parameters.
 
\section{Assessing robustness: Comparing distances to current data}

\begin{figure}[h!]
    \centering
    \includegraphics[width=\textwidth]{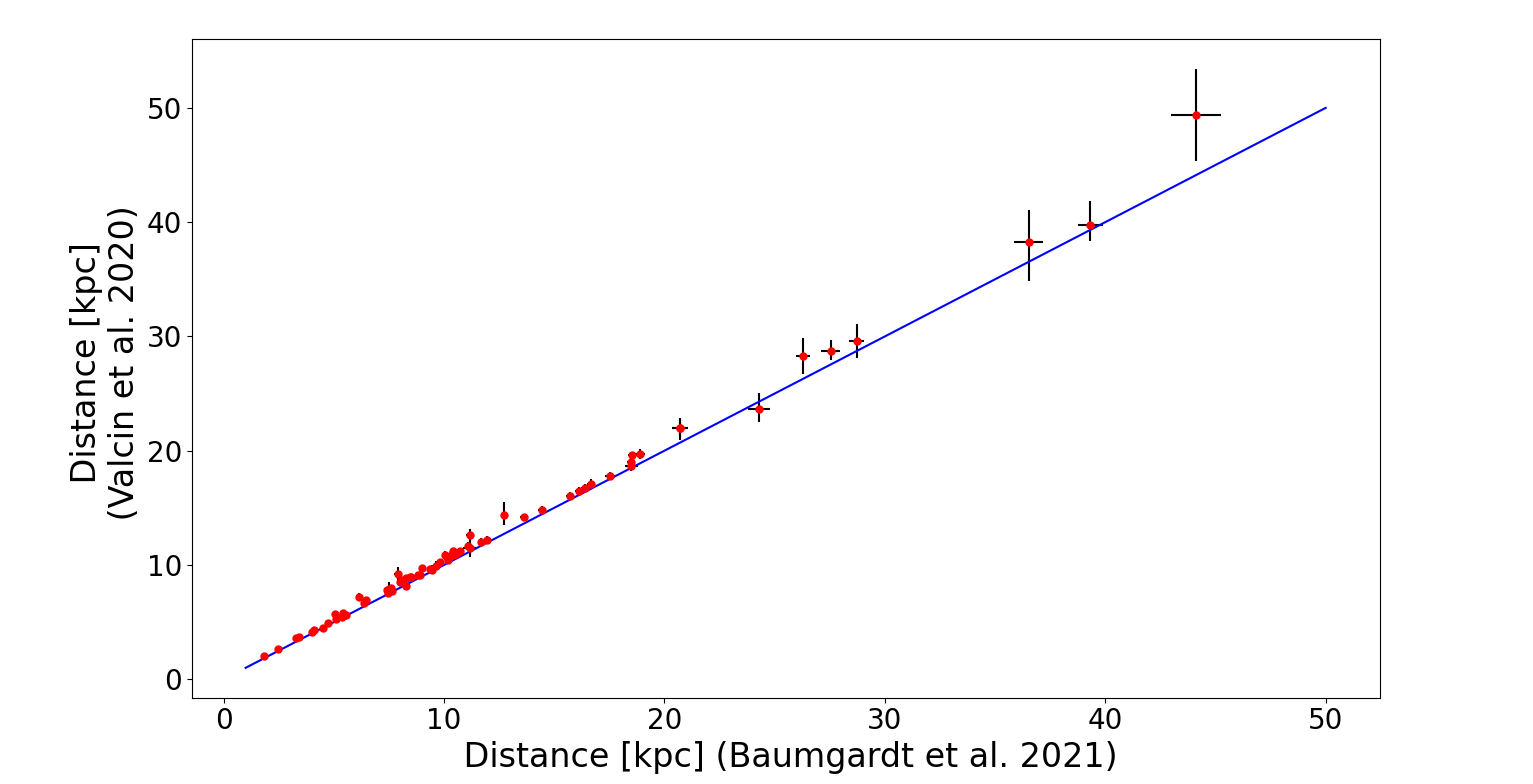}
    \includegraphics[width=\textwidth]{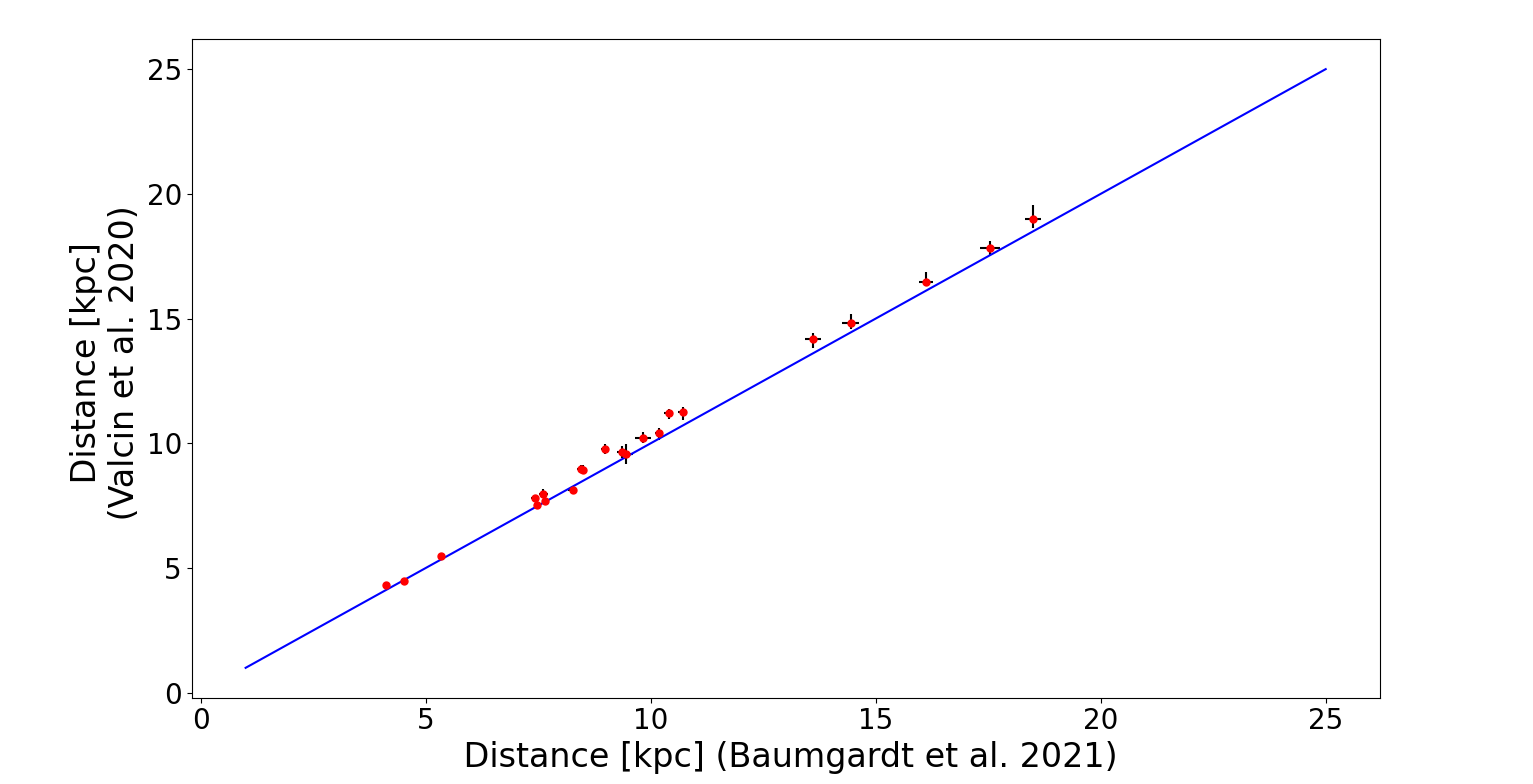}
    \caption{Comparison of distances between Valcin et al. and Baumgardt et al for the full sample of GCs (left) and the 22 of O'Malley et al. (right)}
    \label{fig:comp_gaia}
\end{figure}

As we explained in section \ref{sec:data}, our methodology requires working with the absolute magnitude of the clusters in order to calibrate $\alpha_{\rm LMT}$ for the different samples of metallicity. The conversion from apparent magnitude to absolute magnitude largely depends on the distance modulus or distance of the clusters. As we used the best fits obtained from Valcin et al. \cite{ValcinGC} to correct the magnitudes, it was important to test their robustness against current data (here the third data release of the \texttt{GAIA} catalog \cite{Baumgardt}). We can see in figure \ref{fig:comp_gaia} that the agreement is excellent, especially for the closest clusters which confirms the quality of the overlap of the different clusters within each sample seen in figure \ref{fig:allGC}.

\providecommand{\href}[2]{#2}\begingroup\raggedright\endgroup

\end{document}